\documentclass{aa}
\usepackage{graphicx}
\usepackage{txfonts}

\begin{document}

\title{Magnetic fields and gas in the cluster--influenced spiral\\
galaxy NGC\,4254 
-- II. Structures of magnetic fields
}

\author{Krzysztof T. Chy\.zy}
\institute{Astronomical Observatory, Jagiellonian
University, ul. Orla 171, 30-244 Krak\'ow, Poland
  }

\date{Received / Accepted }

\titlerunning{Magnetic fields and gas in NGC~4254}
\authorrunning{K.T. Chy\.zy}

\abstract
{}
{
The origin of asymmetric radio polarized emission in the Virgo Cluster 
spiral NGC\,4254 is investigated and the influence of cluster environment 
on the properties of magnetic fields is explored. 
}
{Structures of magnetic fields are analyzed with the concept of ``magnetic 
maps'', presenting distributions of different magnetic field components (total, 
regular, and random) over the entire galaxy, free of Faraday rotation and 
projection effects. A number of different physical 
phenomena influencing the magnetic field are modeled analytically and 
confronted with the galaxy's depolarization pattern and distribution 
of magnetic field strength obtained from multifrequency polarimetric radio observations. 
}
{
The study of orientation of intrinsic magnetic field vectors in NGC\,4254 
indicates that their dramatic variation (from $0\degr$ to more than $40\degr$) 
throughout the galaxy cannot arise from the dynamo process alone, but 
must be dominated by effects such as density waves and local gas flows. We 
determine within the galaxy the relation between the strength of 
total magnetic field and the local star-formation rate (SFR) as a power-law with
an index of $+0.18\pm 0.01$. We find the opposite sense of the relation between 
magnetic field regularity and SFR ($-0.32\pm 0.03$), and suggest that it results from efficient 
production of random field with rising turbulence in the regions with 
actively-forming stars. The distribution of Faraday rotation measures in NGC\,4254 indicates 
a perturbed axisymmetrical mean-field dynamo mode or a mixture of axisymmetrical 
and bisymmetrical ones with regular field directed outwards from the disk, which is 
contrary to most observed galaxies. The galaxy's northern magnetic arm, located 
on the upstream side of the local density wave, with regular field strength of 
about $8\,\mu$G and the total one of $17\,\mu$G, much resembles those observed 
in other galaxies. But the magnetic field within two outer arms (shifted 
downstream of a density wave) is much stronger, up to $13\,\mu$G in the regular 
field component and $20\,\mu$G in the total field. 
Our modeling of cluster influence on different magnetic field components 
indicates that within the outer magnetic arms the dynamo-induced magnetic 
fields are modified by stretching and shearing forces rather than by cluster 
ram pressure. Those forces, which are likely triggered by the galaxy's 
gravitational interaction, produce an anisotropic component of the regular 
field and enhance the polarized emission. 
We also show that the magnetic energy within the large interarm regions and 
the galaxy's outskirts exceeds the gas thermal and turbulent 
energy, likely becoming dynamically important.
}
{}

\maketitle

\section{Introduction}
\label{intro}

Disk galaxies in galaxy clusters are particularly affected by the cluster 
environment (Roediger \& Hensler \cite{roediger05}) and can be even entirely transformed from 
one Hubble type to another (Moore et al. \cite{moore96}). Interstellar medium (ISM) within 
a cluster galaxy is altered in a different manner in individual galactic 
regions, as can be seen, e.g., from H$\alpha$ (Koopmann \& Kenney \cite{koopmann04}) 
and \ion{H}{i} (Cayatte et al. \cite{cayatte90}) distributions. The magnetic field in cluster galaxies
as one of the ISM ingredients is also expected to be significantly modified, 
mainly by ram-pressure (compression) of hot intracluster medium (ICM) 
or by tidal (stretching/shearing) forces due to galactic encounters. 
Investigation of influence of such external forces on magnetic field gives an independent 
insight into the galaxy evolution within the cluster (Soida et al. \cite{soida06}).
Exploring magnetic field structures can also allow for determining how tangling 
of magnetic field within star-forming regions alters different magnetic field components, or to 
what extent the gaseous streamlines perturbed in cluster galaxies can tune 
the dynamo-induced galactic magnetic fields (Shukurov \cite{shukurov98}, Elstner \cite{elstner05}).
To trace all those processes locally, we need sensitive radio polarimetric data of at 
least kpc-scale resolution, which are available now only for the Virgo Cluster spirals.

\begin{figure}
\includegraphics[width=8.9cm]{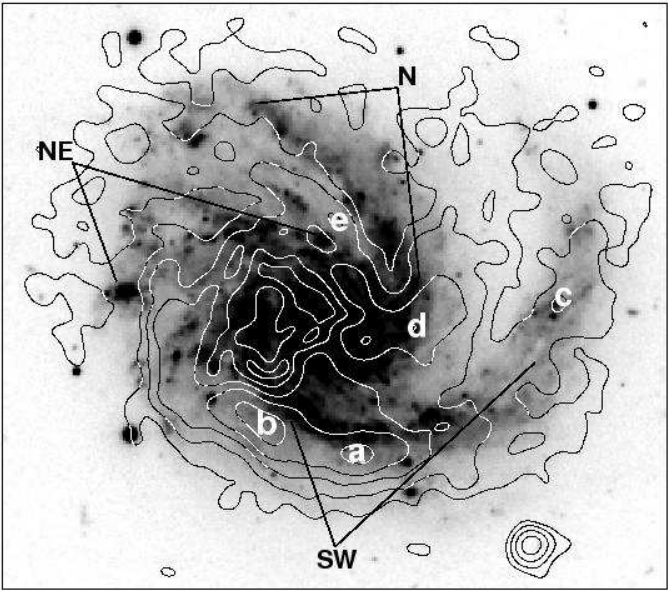}
\caption{Optical DSS B image of NGC\,4254 and polarized intensity
at 4.86\,GHz in $15\arcsec$ resolution in contours. The contours are at $9.7\times (4, 9, 
16, 25, 40)\,\mu$G. Three main optical spiral arms (SW, N and NE) are 
schematically marked. Letters a-e denote positions of contrast derivations in
Sect.~\ref{s:cluster}. 
}
\label{f:intro}
\end{figure}

In this paper we present the first detailed analysis 
of magnetic field, rotation measure, and depolarization patterns of a 
cluster spiral -- the perturbed Virgo Cluster galaxy NGC\,4254. 
In Paper I (Chy\.zy et al. \cite{chyzy07}) we presented our radio polarimetric 
observations (of the VLA of 
NRAO\footnote{National Radio Astronomy Observatory is a facility of National
Science Foundation operated under a cooperative agreement by Associated
Universities, Inc.} and 100-m Effelsberg telescope\footnote{The
100--m telescope at Effelsberg is operated by the Max--Planck--Institut f\"ur
Radioastronomie (MPIfR) on behalf of the Max--Planck--Gesellschaft.})  
and X-rays observations (from XMM-Newton\footnote{
XMM--Newton is an ESA science mission with instruments and contributions directly
funded by ESA Member States and NASA (Jansen et al. \cite{jansen01})}), 
together with the data reduction process, as well as a global outline  
of the galaxy's radio and polarimetric properties in 
relation to other spectral bands. We found that in the radio domain the galaxy 
has an asymmetrical distribution extending to the north, which resembles 
disturbed distributions in the optical, UV, and X-rays bands, dominated 
by a one-arm spiral structure. The polarized intensity maps at 8.46\,GHz 
and 4.86\,GHz show an unusual strong ridge in the southern disk portion, 
outside of a heavy optical spiral arm, with observed magnetic field 
vectors directed along it (Fig.~\ref{f:intro}, see also Fig.~2 in Paper I). 
Similar, but weaker, structures shifted off optical 
features are also visible in other parts of the galaxy disk. Some polarized 
features are strong enough to appear within optical filaments even in the 
highly turbulent spiral arm regions. These mixed magnetic field patterns 
are in sharp contrast to the previous studied grand-design spirals, which 
typically show either interarm or arm-dominated polarized structures (like in 
NGC\,6946 -- Beck \& Hoernes \cite{beckhoernes96} and M\,51 -- Fletcher et al. 
\cite{fletcher04b}, respectively). Moreover, the star-formation rate (SFR) derived 
from radio thermal emission is higher in NGC\,4254 than in 
other galaxies, indicating some effects of an external (cluster) agent.

The pattern of magnetic fields discovered in NGC\,4254 is investigated in the 
present paper in a detailed manner. In Sect.~\ref{s:rm}, we present the 
Faraday rotation measure and depolarization maps and derive 
a Faraday-free intrinsic structure of magnetic field. Next, we introduce 
``magnetic maps'' (Sect.~\ref{s:magmaps}) -- a novel concept of presenting 
magnetic field strength in different magnetic field components in the form of 
distributions across the entire galaxy, free of Faraday rotation and projection 
effects. In Sect.~\ref{s:orient}, we study the orientation (pitch angle) of 
magnetic field vectors within structures of optical arms and gas filaments. 
We investigate regularity of magnetic field over 
regions of different star formation rates (Sect.~\ref{s:sfr}) and model 
various Faraday depolarization effects (Sect.~\ref{s:depo}). The impact of 
external compression and stretching effects on the different components of 
galactic magnetic field is analytically modeled in Sect.~\ref{s:ani} 
and finally discussed in Sect.~\ref{s:cluster}. 

\section{Results}
\subsection{Distribution of Faraday rotation and depolarization}
\label{s:rm}

\begin{figure*} 
\includegraphics[width=8.9cm]{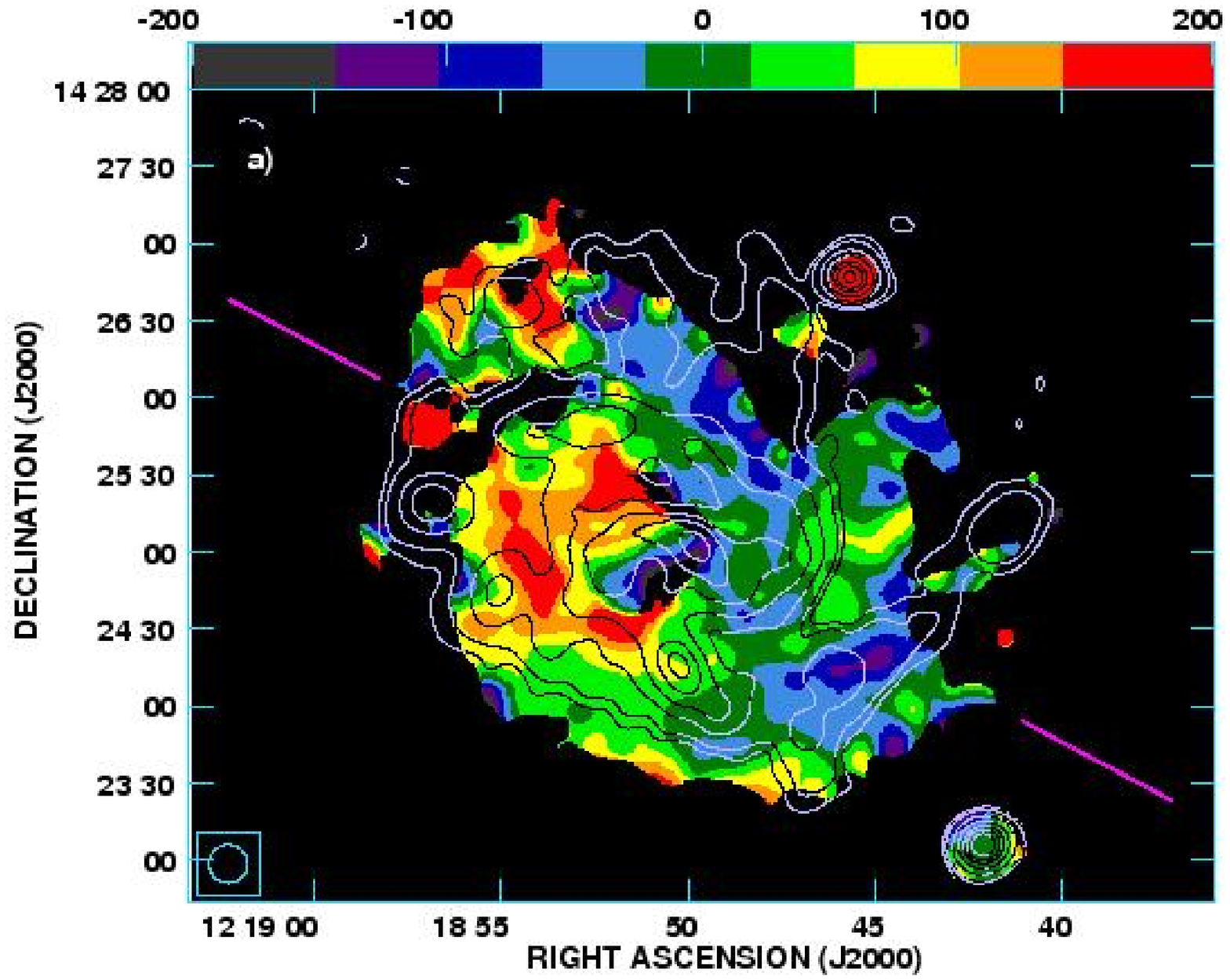}
\includegraphics[width=8.9cm]{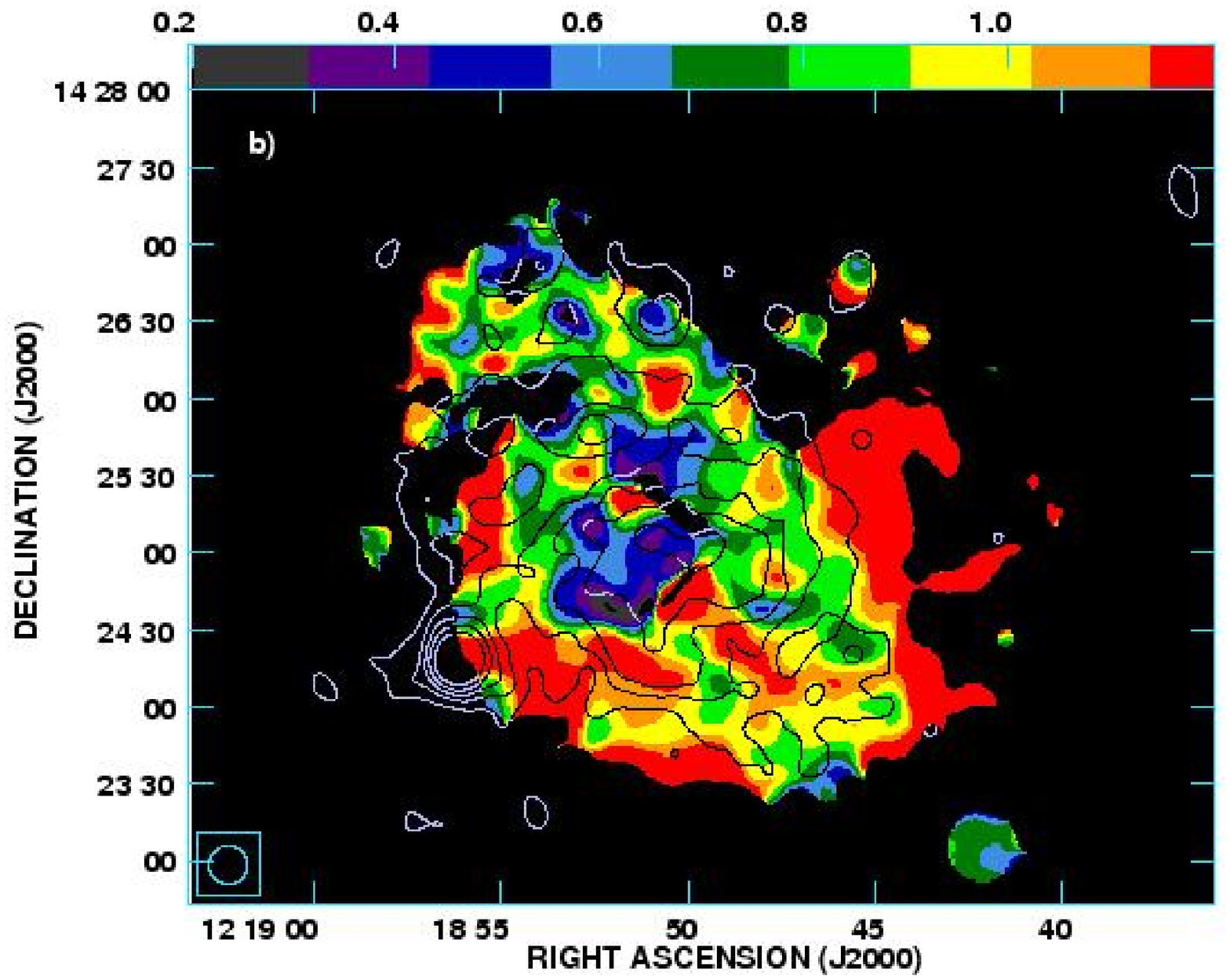}
\caption{
a) Faraday rotation measure distribution in NGC\,4254 in $15\arcsec$ 
resolution in colors, computed between 8.46\,GHz and 4.86\,GHz with contours 
of total radio intensity at 4.86\,GHz. The galaxy major axis is marked. b) 
Faraday depolarization distribution in colors and contours of soft X-ray 
emission from our XMM-Newton observations.
} 
\label{f:rm}
\end{figure*}

For the purpose of analyzing magnetic field in NGC\,4254 we use the radio 
polarimetric observations described in detail in Paper I. The 
observed polarized component of synchrotron emission results from 
the regular (ordered) component of magnetic field. It can contain a mixture 
of coherent (unidirectional) and anisotropic (incoherent) magnetic 
fields. Only the coherent field contributes to the observed Faraday 
rotation because the contributions from an anisotropic field with 
vectors of opposite directions cancel when averaged over the telescope beam. 
To construct the Faraday rotation measure ($RM$) map we take the 
distributions of polarization angles at 4.86 and 8.46\,GHz in 
$15\arcsec$ resolution. The observed Faraday rotation also includes 
the contribution from the Galaxy's foreground rotation measure. 
The spiral NGC\,4254 lies at $270.4\degr$ Galactic longitude and $75.2\degr$ 
Galactic latitude, where Johnston-Hollitt et al. (\cite{johnston04}) 
show only low values of foreground $RM$, changing sign from place 
to place with absolute values $<25$\,rad\,m$^{-2}$. The average value of $RM$ 
estimated from our data over the entire galaxy is also small: about  
+15\,rad\,m$^{-2}$. Some reliable measure of foreground $RM$ can be obtained 
from a polarized background source visible to the SW from NGC\,4254 
(${\rm RA}_{2000}=12^{\rm h}18^{\rm m}42\fs1$, ${\rm Dec}_{2000}=+14\degr 
23\arcmin 07\arcsec$, Fig~\ref{f:rm}a). 
Within the uncertainties in measurement of about $3\degr$, the source 
shows no difference in polarization position angle between 8.46 and 4.86\,GHz, 
thus suggesting a low RM ($< 20$\,rad\,m$^{-2}$). Between 4.86 and 1.43\,GHz we 
measured the position angle difference of $5.5\degr$, which provides very 
small $RM$ of 2.5\,rad\,m$^{-2}$. Therefore, we do not apply any correction 
to the data for the foreground Faraday rotation.

The $RM$ map for NGC\,4254 is presented in Fig.~\ref{f:rm}a together 
with contours of total radio emission. Over many galactical regions the 
absolute values of $RM$ are typically small, of the order of 
50-70\,rad\,m$^{-2}$ and exceed 100\,rad\,m$^{-2}$ only within small regions. 
The variation of $RM$ over the disk can reflect various physical processes 
(see Sect.~\ref{s:rm}) and local changes of magnetic field 
strength as well as its orientation with respect to the observer. In general, 
the western part of the galaxy has negative $RM$ values, while the 
eastern part has positive ones that are slightly 
larger in the absolute sense. This demonstrates a substantial coherence of the 
sign of $RM$ and hence of the magnetic field orientation. Such coherence of 
magnetic field strongly suggests a global galactic dynamo in action (Beck et al. 
\cite{beck96}, Widrow \cite{widrow02}).

Close to the center of NGC\,4254 there are two sudden jumps of $RM$ oriented
at 90$\degr$ to each other (${\rm RA}=12^{\rm h}18^{\rm m}51\fs0$, ${\rm Dec}
=+14\degr 24\arcmin 31\arcsec$; and ${\rm RA}=12^{\rm h}18^{\rm m}52\fs0$, 
${\rm Dec}=+14\degr 25\arcmin 20\arcsec$). In these places, our map keeps track 
of $RM$ values up to about $\pm 400\,$rad\,m$^{-2}$. Such jumps usually 
mean that the internal Faraday rotation angle exceeds $90\degr$ and indicate a 
Faraday-thick regime (Sokoloff et al. \cite{sokoloff98}). The jumps do not 
correspond to any feature in the radio or \ion{H}{i} emission. However, the 
southern jump is close to a CO extension going radially outwards from the 
nuclear ring (see Fig.~3 in Sofue et al. \cite{sofue}).

The region of the SW optical arm where the strongest polarized emission 
in NGC\,4254 is observed shows a rather small amount of Faraday rotation: 
around its eastern polarized peak (marked as "a" in Fig.~\ref{f:intro}) the 
$RM$ is about $-30\pm 7$\,rad\,m$^{-2}$. Around the western peak (marked as "b" in 
Fig.~\ref{f:intro}) the $RM$ has the opposite sign and a value of about 
$+20\pm 6$\,rad\,m$^{-2}$. This may indicate a change of magnetic field 
orientation with respect to the observer. In fact, the two peaks are on both 
sides of the galaxy's minor axis (see Fig.~\ref{f:rm}a with marked the galaxy's 
major axis at the position angle of $68\degr$ envisaged from the \ion{H}{i} 
gas kinematics, Phookun at al. \cite{phookun93}).

Faraday rotation effects not only change the orientation of actual 
magnetic field vectors but also modify (typically reduce) the observed degree of 
polarization. We measure the wavelength-dependent depolarization $DP$ as the ratio 
of the degree of polarization ($p$) of nonthermal emission at 4.86\,GHz 
(6.2\,cm) and 8.46\,GHz (3.5\,cm):
\begin{equation}
DP=p_{4.86}/p_{8.46}.
\end{equation}
We separated the nonthermal emission at both frequencies from the total radio 
intensity with our observations at 1.43\,GHz, 4.86\,GHz, and 8.46\,GHz, assuming 
a constant nonthermal spectral index $\alpha_{\rm nth}=1.0$ (cf. Paper I). 
The map of depolarization of NGC\,4254 presented in Figure~\ref{f:rm}b 
exhibits globally an asymmetric pattern, as in the case of the $RM$ map. The 
southern and western parts of the galaxy show very little depolarization 
($DP>0.9$). This is in agreement with substantially-polarized emission observed here 
at a low frequency of 1.43\,GHz (Paper I), which is an independent indication of 
Faraday-thin regime. 

In the eastern galaxy part depolarization is stronger, with $DP$ values of 
about 0.6-0.8. The strongest depolarization effects occur at the nucleus, 
where locally $DP\approx 0.2$, and around $RM$ jumps, where $DP<0.2$. These 
low $DP$ values must have been produced by a strong Faraday rotation within 
the emitting region (Sokoloff et al. \cite{sokoloff98} and Sect.~\ref{s:depo}). 
All those central regions are actually associated with strong H$\alpha$ and 
hot gas (soft X-ray) emissions. As expected, they also show rather weak 
polarized intensity at 1.43\,GHz (less than 0.03\,mJy) and a weak degree 
of polarization ($<1\%$).

We attempt to identify main processes to affect polarization in NGC\,4254 by
performing a mutual comparison between $DP$, $RM$, and radio thermal emission 
in 136 beam-independent regions within the galaxy. There is no statistically 
important global correlation of the thermal gas emission with the $DP$ or 
$RM$ values: the Pearson correlation coefficient $r$ is around $-0.10$. 
The relation of $DP$ with $RM$ is also weak ($r=-0.02$). Thus,  
no single process associated with the galactic thermal gas or the regular 
magnetic field is capable of accounting for the $DP$ distribution. Therefore, a more 
detailed analysis in separate regions in the galaxy is needed (see 
Sect.~\ref{s:depo}).

\begin{figure} 
\includegraphics[width=8.9cm]{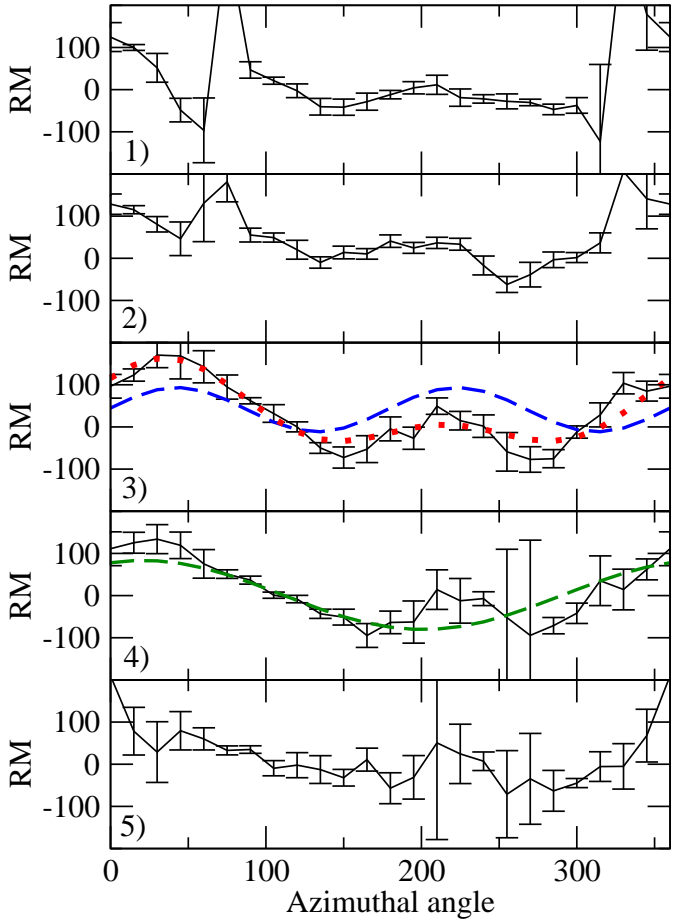}
\caption{
Distribution of rotation measures between 8.46 and 4.86\,GHz along the 
azimuthal angle in 5 different rings of NGC\,4254. The rings of 1.2\,kpc width 
start from $30\arcsec$ (2.4\,kpc) distance from the galaxy center (ring 1, the 
most upper one). The last ring (bottom) ends at distance of 8.5\,kpc. The 
best-fitted dynamo modes are also shown: for the ring 3 the bisymmetric m=1 mode 
(dashed) and superposition of axisymmetric and bisymmetric m=0/1 modes 
(dotted); and for ring 4 the axisymmetric m=0 mode (dashed), see Table~\ref{t:modes}.
} 
\label{f:rmsect}
\end{figure}

\begin{figure}[t] 
\includegraphics[width=9.0cm]{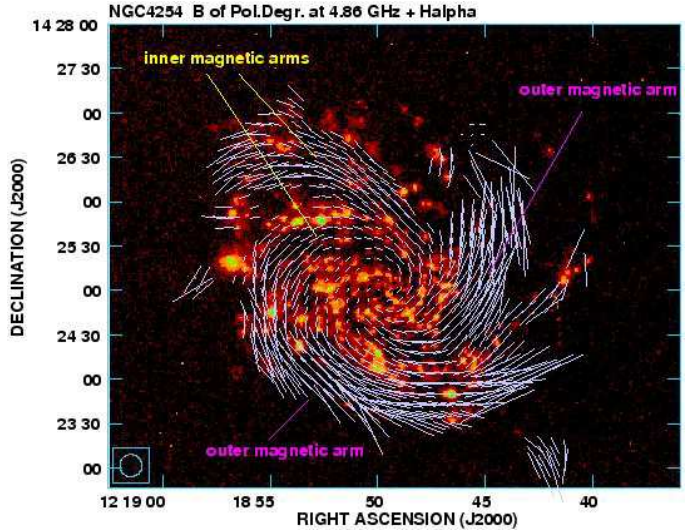}
\caption{
Actual (Faraday-free) structure of magnetic field vectors in NGC\,4254 obtained 
from the combined VLA and single-dish (Effelsberg) polarized data at 8.46\,GHz 
and 4.86\,GHz. The lengths of vectors are proportional to the degree of 
polarization at 4.86\,GHz. The regions with inner and outer magnetic arms are 
designated. Overlaid (in colors) is the H$\alpha$ image (from Knapen et al. \cite{knapen04}).
} 
\label{f:true}
\end{figure}

\subsection{Genuine structure of magnetic field}
\label{s:genuine}

The same sign of $RM$ values over large regions in NGC\,4254 (Sect.~\ref{s:rm}, 
Fig.~\ref{f:rm}a) most probably originates from a dynamo-generated magnetic field. 
In order to recognize a dominant dynamo mode in action we determined a 
distribution of $RM$ within the galactic plane along rings around the disk 
center. The rings of radial width of 1.2\,kpc (corresponding to the beam 
size of 15\arcsec) were split along azimuthal angle into sectors of $15\degr$ width. 
For each sector we calculated the average rotation measure and we present the 
distribution of $RM$ in five rings in Fig.~\ref{f:rmsect}. 
Then, in the most characteristic rings 3 and 4, encompassing 
the southern polarized ridge, we fit different periodic variations 
of $RM$ resulting from dynamo modes and their superpositions. The best fits 
are presented in Table~\ref{t:modes} and Fig.~\ref{f:rmsect}. The $RM$ distributions 
partly resemble single-periodic variations, thus the axisymmetric m=0 mode. 
The fitted phase shifts also correspond well to the observed magnetic pitch 
angles (about $20\degr$) in the first and the second quadrant of the azimuthal angle. 
However, a strong disturbance breaks the dynamo symmetry in the third quadrant 
(the azimuth in the range of $150\degr$ -- $270\degr$). The $RM$ instead of 
achieving, as predicted by the model, a negative minimum of about 
$-200\,\mathrm{rad}\,\mathrm{m}^{-2}$ manifests a more complicated pattern 
involving a local field reversal. This deviation corresponds 
to a strongly-polarized large interarm region between SW/N optical arms, with 
a quite smooth magnetic 
field pattern. Similarly, the bisymmetric (m=1) dynamo mode alone cannot account for 
the $RM$ distribution as it yields too strong a maximum at azimuth of $210\degr$ and 
too weak one at $40\degr$. However, adding it as a secondary component to the 
axisymmetric mode (with roughly half of its amplitude) significantly improves the 
fits. Higher dynamo modes and other mixed modes do not provide any better fits. 
Surprisingly, it is also true for a quadrupolar (m=2) mode and a 
mixture m=0/2 modes, which were suggested to explain the phenomena of 
magnetic arms in NGC\,6946 (Beck \cite{beck07}, Rohde et al. 
\cite{rohde99}). 

\begin{table}[t]
\caption[]{Parameters of the single (m=0) double periodic (m=1) cosine 
functions and their superposition (m=0/1) fitted to the rotation 
measure variations in the two rings of NGC\,4254 (cf. Fig.~\ref{f:rmsect}). 
Amplitude, phase shift, background offset, and the Theil~U coefficient of 
the quality of fit are given. 
}
\begin{center}
\begin{tabular}{lrrrr}
\hline
\hline
Dynamo mode  & ampl.  &  ph. shift & offset & Theil U$^a$ \\
\hline
RING 3\\
m=0          & 79    & 25 & 23 & 0.58 \\
m=1          & 53    & 43 & 41 & 0.83 \\
m=0/1        & 79/51 & 34 & 33 & 0.34 \\
\hline
RING 4\\
m=0          & 82    & 21 &  2 & 0.57 \\
m=1          & 48    & 48 & 17 & 0.94 \\
m=0/1        & 71/36 & 33 &  9 & 0.36 \\
\hline \\
\label{t:dynamo}
\end{tabular}
\end{center}
$^a$~The Theil~U coefficient is limited to a range from 
zero to one (Theil \cite{theil72}). Zero means a perfect fit.
\label{t:modes}
\end{table}

Using the $RM$ and \ion{H}{i} data we can estimate the preferred radial 
direction of magnetic field, e.g., inward or outward of the galactic center (cf. 
Krause \& Beck \cite{krause98}). 
The kinematical properties of the spiral structure in 
NGC\,4254 can provide information on the sense of the galaxy's rotation. As 
argued by Phookun at al. (\cite{phookun93}), the spiral arms 
in NGC\,4254 are generally of trailing type, although a small fraction of stars rotate 
in reverse. From the \ion{H}{i} velocity field we know that the western 
part of the galaxy is in approaching motion, thus the southern part of the disk is the 
nearest side. As both the radial velocity and $RM$ have the same sign on both ends 
of the galaxy's major axis (in particular, magnetic fields are directed outwards from the 
observer in the western part), we conclude that the regular (coherent) field in 
NGC\,4254 is oriented outwards from the disk center, which is contrary to most observed 
galaxies (Krause \& Beck \cite{krause98}). The outwards directed field was observed to 
date only in the disk of M\,51. Thus, in both galaxies, outwards fields could possibly 
arise from the external interaction with a companion. 

With the distribution of $RM$ available at hand, we can reconstruct the intrinsic 
position angle of the magnetic field over the whole galaxy, free of Faraday rotation. 
We assume a simple quadratic relation of $RM$ to wavelength, which should be
an appropriate approximation for the high frequency data used. The actual 
magnetic field structure thus corrected is presented in Fig.~\ref{f:true} 
with vector lengths proportional to the degree of polarization at 
4.86\,GHz. The magnetic structure is smooth and of spiral pattern, even 
within the galaxy inner $1\arcmin$ region, where the optical (H$\alpha$) spiral 
structure is unclear.

The visible magnetic arms in Fig.~\ref{f:true} are mainly interlaced with 
optical (H$\alpha$) ones. However, we distinguish `inner' magnetic arms, 
which are displaced off the optical ones to their inner sides and the 
`outer arms', which are shifted off outwards, to the downstream side 
of a density wave. In the next section, we 
show that both kinds of magnetic arms differ in magnetic properties, which
must have been due to some differences regarding their origin. In general, 
the pattern of magnetic arms in NGC\,4254 apparently follow the structure of 
local optical features, which indicates 
a substantial influence of density waves on magnetic pitch angle 
(see Sect.~\ref{s:orient} and \ref{s:cluster}). 

\subsection{Magnetic maps}
\label{s:magmaps}

\begin{figure*} [t]
\centering
\includegraphics[width=18.5cm]{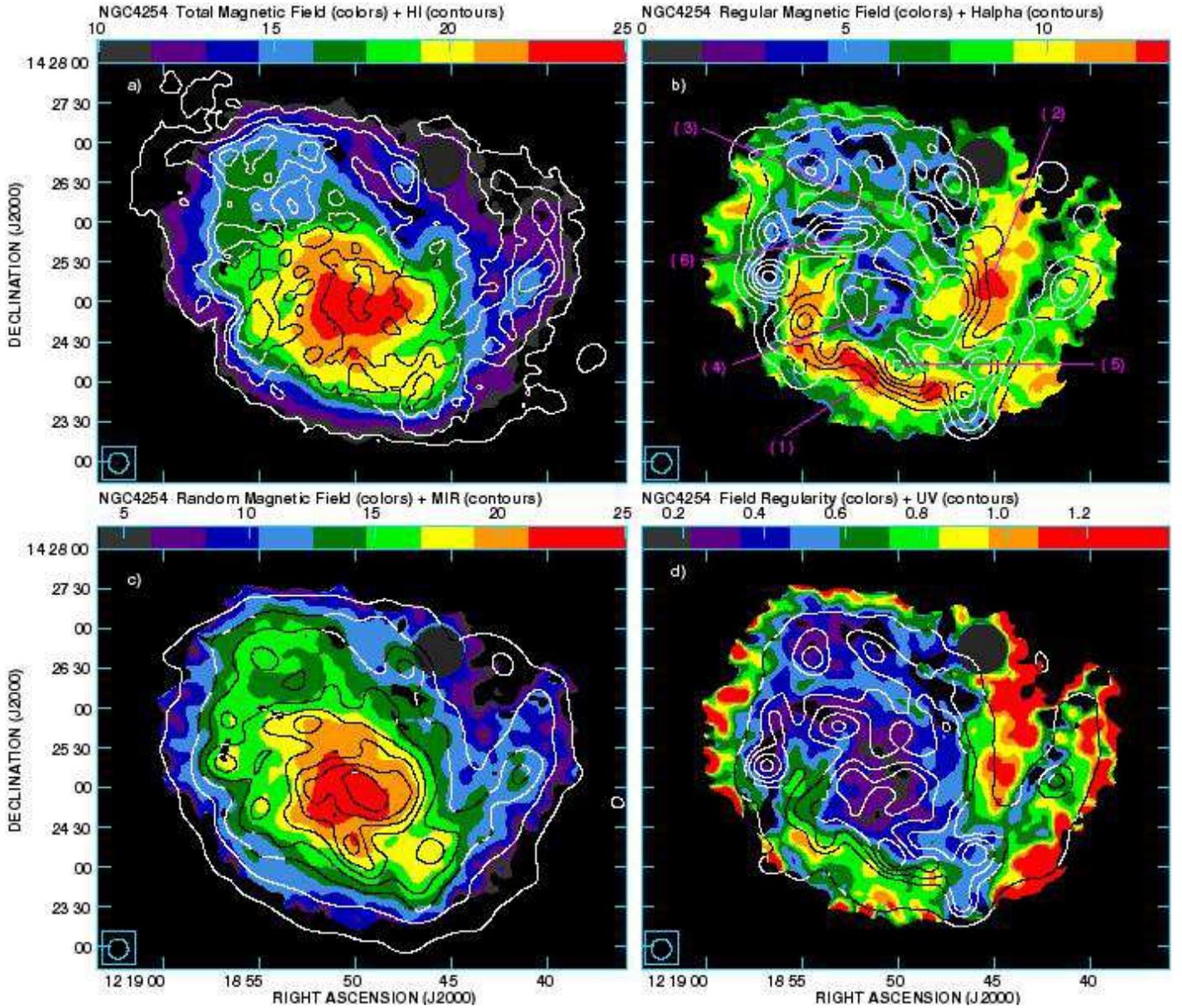}
\caption{Magnetic maps for NGC\,4254 with a resolution of $15\arcsec$ (in colors): total (a), regular (b), 
random (c) magnetic field strength, and field regularity 
(d) with contours of \ion{H}{i}, H$\alpha$, infrared (24\,$\mu$m), and UV emission, 
respectively. The method of the maps construction is given in 
(Sect.~\ref{s:magmaps}). 
}
\label{f:magmaps}
\end{figure*}

Currently, magnetic properties of galaxies are most often presented using the field 
strength estimated in a small number of individual regions within a galaxy or as the average 
for the whole disk. In the case of complicated magnetic field structures as, e.g., 
in the Antennae system of interacting galaxies proper interpretation of polarimetric 
data requires a comparison of field strength in the regular and random components in 
many different regions over the entire system (Chy\.zy \& Beck \cite{chyzy04}). The 
magnetic field in the perturbed spiral NGC\,4254 is quite complicated too, 
showing unusual polarization features and $RM$ pattern, different kind of magnetic arms 
in various parts of the galaxy, and thus requires a similar mode of analysis. 
We, therefore, propose here to present for the first time galactic magnetic field in a form 
of complete ``magnetic maps'' that show the distributions of strengths of
intrinsic magnetic field components in all available regions in the galaxy disk, 
with the galaxy projection and Faraday rotation effects removed.

\begin{table*}[t]
\caption[]{
The intrinsic magnetic field strength of total $B_{\rm{tot}}$ and regular $B_{\rm{reg}}$ 
components as well as the field regularity in different regions of NGC\,4254, 
denoted in Fig.~\ref{f:magmaps} (top-right). The given uncertainties express random noise 
in the maps used. The errors, including possible systematic uncertainties due 
to applied assumptions, can reach 25\% of the values obtained. The presented 
magnetic fields strengths are taken directly from magnetic maps described 
in (Sect.~\ref{s:magmaps}).
}
\begin{center}
\begin{tabular}{lrrr}
\hline
\hline
                       &  $B_{\mathrm{tot}}$         &    $B_{\mathrm{reg}}$  &  ${B_{\mathrm{reg}}}/{B_{\mathrm{ran}}}$   \\
                        & [$\mu$G]         &  [$\mu$G]     &                              \\
\hline
Total                   &   $16 \pm 1$     & $7 \pm 0.2$   &   n.a.                      \\
1) Southern outer magnetic arm &   $19.7 \pm 0.5$     & $12.6 \pm 0.3$  & $0.83 \pm 0.03$ \\
2) Western outer magnetic arm  &   $18.9 \pm 0.8$     & $12.3 \pm 0.4$  & $0.86 \pm 0.09$ \\
3) Northern inner magnetic arm &   $16.9 \pm 0.2$     & $ 7.8 \pm 0.3$  & $0.52 \pm 0.02$ \\
4) Central part                &   $24.3 \pm 0.5$     & $ 4.8 \pm 0.7$  & $0.20 \pm 0.03$ \\
5) SW optical spiral arm       &   $22.2 \pm 0.3$     & $ 7.4 \pm 0.2$  & $0.35 \pm 0.02$ \\
6) NE optical spiral arm       &   $20.3 \pm 0.7$     & $ 6.5 \pm 0.8$  & $0.34 \pm 0.05$ \\
\hline\\
\label{t:mf}
\end{tabular}
\end{center}
\end{table*}

The strength of {\em intrinsic} regular 
component of magnetic field $B_{\rm{reg}}$, corrected for galaxy 
inclination and varying angle between the line of sight and the field line, is 
derived from the observed radio polarized intensity. We assume that the regular 
magnetic field lies only in the galaxy plane, which permits derivation 
of the intrinsic (in the disk plane) magnetic field position angle from the 
derived sky-projected (but Faraday rotation free) polarization position angle 
(Sect.~\ref{s:rm}) and the galaxy orientation. Following Phookun et al. 
(\cite{phookun93}) we adopt values of $42\degr$ and $68\degr$, respectively, 
for the galaxy inclination and the major axis position angle. The intrinsic 
(de-projected) total magnetic field strength $B_{\rm{tot}}$ is calculated 
through a similar procedure using, in this case, the total nonthermal 
(synchrotron) intensity. The random magnetic field component $B_{\rm{ran}}$ 
is derived by subtracting the reconstructed intrinsic regular field from the 
total field. 

In our calculations, we apply the energy ratio $k=100$ of relativistic 
protons and electrons, and a 300\,MeV cutoff in the cosmic rays (CR) 
proton spectrum (e.g., Ehle \& Beck \cite{ehle93}; Chy\.zy \& Beck 
\cite{chyzy04}). We assume the equipartition energy condition between CRs 
and magnetic fields and apply a constant nonthermal spectral index 
$\alpha_{\rm nth}=1.0$ (c.f. Paper I). The magnetic field is supposed to be 
located within the plane of the galaxy disk (without a vertical 
component) of about 1\,kpc unprojected thickness. The whole 
described procedure of magnetic maps' derivation was implemented as a task in the 
AIPS\footnote{Astronomical Imaging Processing System of the National 
Radio Astronomy Observatory}. The largest uncertainties in the obtained 
magnetic field strengths are most likely to be the systematic ones due to 
the assumed above values for the unknown parameters, rather than random 
errors due to uncertainties in observed total and polarized radio 
intensities (see below). 

The first magnetic map of NGC\,4254 presented in Fig.~\ref{f:magmaps}a 
shows the distribution of {\em total} magnetic field strength with contours of 
\ion{H}{i} intensity for reference sake. In the central disk region the strength reaches 
25\,$\mu$G, while decreasing to about 10\,$\mu$G at the outer galaxy parts (see 
also Table~\ref{t:mf}). The strong localized signal in the disk northern 
periphery (RA=$12^{\rm h}18^{\rm m}46^{\rm s}$,
Dec=$14\degr26\arcmin 45\arcsec$) coming from a confusing 
background source is masked in this and following magnetic maps. 
The mean strength of total magnetic field over the 
whole galaxy (without the background source) is $16\pm1$~$\mu$G, which is in agreement with the 
low-resolution estimation by Soida et al. (\cite{soida96}). This value is 
higher than typical ones (of about $10\,\mu$G) found in a large sample of 
galaxies (Beck et al. \cite{beck96}) but closer to the mean field strength 
in the disks of the Antennae galaxies (about $20\,\mu$G, Chy\.zy \& Beck 
\cite{chyzy04}). As can be seen from Fig.~\ref{f:magmaps}a (and 
Table~\ref{t:mf}), the sites of star formation 
and effective production of CRs in optical spiral arms possess a strong magnetic 
field of about $16-22\,\mu{\rm G}$. In the interarm regions, the total field 
weakens typically to $14-17\,\mu{\rm G}$. We mention that in some analyses of 
galaxies of Hubble type similar to NGC\,4254 a disk thickness of 0.5\,kpc is 
assumed. If, by way of comparison, this value is applied to NGC\,4254, the 
corresponding total magnetic field in the optical arms rises to 
$19-25\,\mu{\rm G}$ and to $30\,\mu{\rm G}$ in the disk core.

The second magnetic map of intrinsic {\em regular} field is presented 
in Fig.~\ref{f:magmaps}b with contours of H$\alpha$ emission. The 
typical strength of regular field in the N and NE optical arms is about 
$5\,\mu$G. In the inner magnetic arm (region 3) it rises up to 
$7.5-8\,\mu$G. In the SW optical spiral arm, the regular field reaches
$6-7\,\mu$G, while outside it, in the southern outer magnetic arm (in 
the strongly-polarized ridge, region 1) it is enhanced up to $13\,\mu$G. The large 
interarm region between SW and N optical arms maintains over a large area, a strong 
regular field of about $10\,\mu$G, stronger than in the northern inner magnetic 
arm. In the western outer magnetic arm (region 2), it reaches a value even as 
strong as the regular field in the southern polarized ridge: $12-13\,\mu$G. 

The distribution of intrinsic regular field shows a global asymmetry in the 
SW-NE direction that was partly hidden due to projection effects in the 
distribution of polarized intensity, which shows just a strong N-S asymmetry 
(Fig.~\ref{f:intro}, Paper I). The strong regular field, manifested not only in 
the southern galactic outskirts, but also in the western outer magnetic arm 
(region 2) {\em inside} the galactic disk, must have profound consequences. 
This introduces a question of the origin of the field in both the regions, and 
their possible connection.

A map of the turbulent component of magnetic field 
(Fig.~\ref{f:magmaps}c) shows an excellent 
correlation with the dust emission (in contours) 
observed in mid-infrared (MIR) by Spitzer (see Paper I). This 
can indicate a strong production of turbulent field or a strong tangling 
(de Avillez \& Breitschwerdt \cite{deavillez05}) in 
sites of star-forming regions revealed without extinction by the MIR 
emission. Surprisingly, in the place of strongly-increased regular field 
in the southern outer magnetic arm no such 
enhancement is visible in the random field. We model this behavior of 
different magnetic field components in Sect.~\ref{s:ani}. 

The ratio of derived (projection free) regular to random magnetic field 
components -- the field regularity ${B_{\mathrm{reg}}}/{B_{\mathrm{ran}}}$ 
-- is a useful measure of the net 
production of regular field independent of magnetic field strength, 
as well as a sensitive tracer of local processes in magnetized plasma. We 
present the distribution of field regularity as yet another ``magnetic 
map'' in Fig.~\ref{f:magmaps}d, together with the contours 
of UV emission associated with population of young stars and the regions of 
most turbulent ISM. In fact, along the three main optical spiral arms and in 
the galaxy center the field regularity is low ($0.2-0.5$). Both the magnetic outer 
arms (region 1 and 2) have an exceptionally high-field regularity of about 
0.8, in contrast to a typical value of 0.5 in other interarm regions 
(including region 3). As it could be expected, the regularity is also larger when 
going outside turbulent regions into the galaxy outskirts. Hence, such a 
highly-regular magnetic field in the galactic halo is likely to enter the 
Virgo Cluster's medium. 

With the magnetic maps at hand, we also looked at where the magnetic energy 
dominates the total energy in the thermal gas component. We derived the 
distribution of the latter quantity from the emission measure estimated from 
radio thermal emission (Paper I). The magnetic energy exceeds 
the thermal one across almost the entire galaxy (except for just a few regions 
of very intensive star formation). Moreover, it exceeds the turbulent gas energy 
(estimated from the dispersion in the \ion{H}{i} velocity field, Paper I) in 
the southern polarized ridge, in the large western interarm region, 
and in the whole polarized galactic outskirts visible at 4.86\,GHz and 
1.4\,GHz. Thus, in those regions the magnetic field can be even dynamically 
important (as in NGC\,6946; Beck \cite{beck07}).

In the above estimations of the magnetic field strength, we assumed a constant 
synchrotron spectral index $\alpha_{\rm nth}=1.0$ across the galaxy, while 
it may attain smaller values in star-forming regions and larger ones in the 
galaxy's outskirts. Such variations, if known, could easily be incorporated 
in the procedures calculating the magnetic maps. Fortunately, the strength of 
the magnetic field only weakly depends on $\alpha_{\rm nth}$. For example, 
we performed an additional separation of thermal and nonthermal radio emission 
assuming $\alpha_{\rm nth}=0.8$, which is more suitable for spiral arms. Then, 
we derived the magnetic maps with the new $\alpha_{\rm nth}$ and compared 
them with the previous ones. In spiral arms and in the galactic centre the 
differences are up to about $2\,\mu$G. In particular, in the optically-bright 
region in the NE spiral arm at RA=$12^{\rm h}18^{\rm m}53^{\rm s}$, 
Dec=$14\degr25\arcmin 51\arcsec$ the total field strength is smaller by 
$1.3\,\mu$G, the regular one by $0.58\,\mu$G, and the random one by $1.4\,\mu$G, 
giving the field regularity of 0.22, instead of 0.30 before. 
Due to energy losses of the CR electrons, the field strength can be 
underestimated in the outer part of the galaxy due to a higher 
ratio $k$ of proton to electron energy. We estimated that effect by setting 
$k$ to 200. Differences to the original maps are in this case up to about $3\,\mu$G. 
In the tip of western outer magnetic arm at 
RA=$12^{\rm h}18^{\rm m}44^{\rm s}$, Dec=$14\degr26\arcmin 21\arcsec$ 
the total field is stronger now by $2.4\,\mu$G, the regular one by $2.1\,\mu$G, 
and the random one by $0.6\,\mu$G. The field regularity rose from 1.25 
to 1.41. Such variations of parameters across NGC\,4254, as discussed above, 
would not change conclusions of this paper.

\section{Discussion}
\label{s:discussion}

\begin{figure*}
\includegraphics[width=18.5cm]{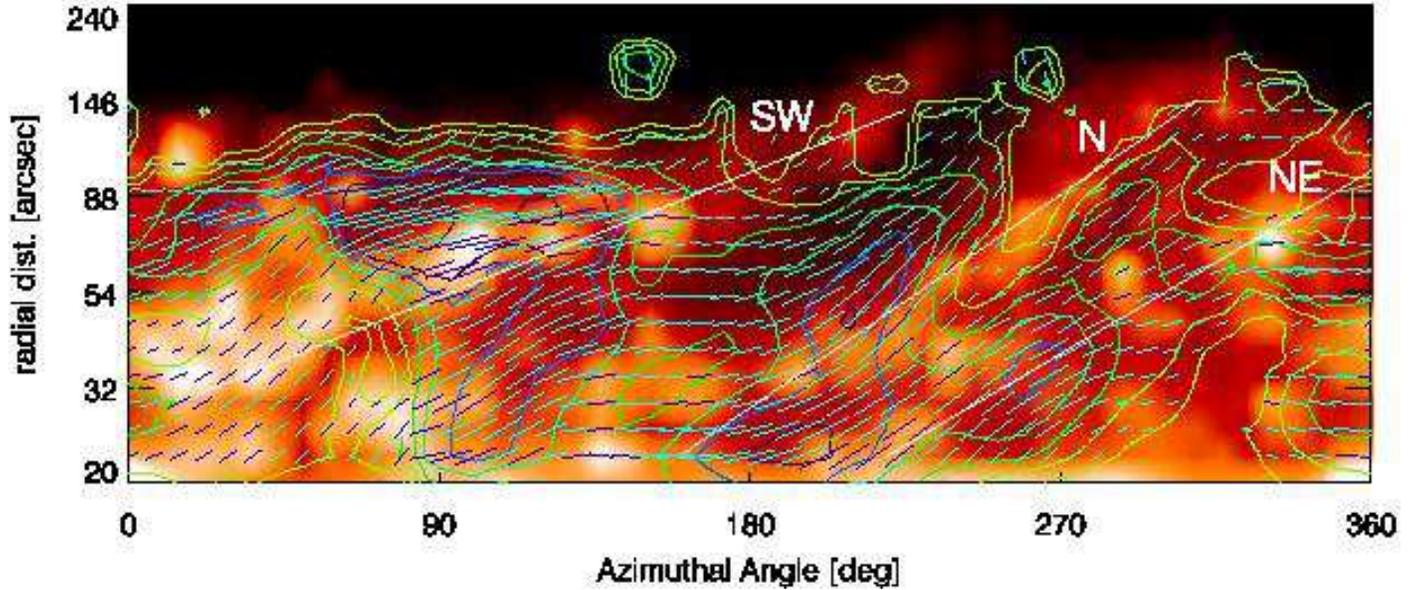}
\caption{A phase diagram of the regular component of magnetic fields in
dependence of natural logarithm of distance and galactocentric azimuthal
angle, as measured from the northern tip of the major axis counterclockwise.
The vectors are proportional to the polarized intensity at 4.86\,GHz, which is
also presented in contours. In the frame of this representation, a logarithmic
spiral arm would be a straight line inclined by the pitch angle. The galactic 
spiral structure is shown by the distribution of UV emission (in color) with
additionally marked (the straight lines) approximate orientations of the main 
three optical spiral arms. All the presented values were averaged in sectors of 
$7\degr$ width in the azimuth.
}
\label{f:orient}
\end{figure*}
 
The intrinsic magnetic field vectors in NGC\,4254 seem to follow the perturbed 
optical spiral pattern (Fig~\ref{f:true}). This suggests that apart from the 
dynamo action, an additional influence of gas flows is involved. 
The $RM$ distribution also reveals a perturbed axisymmetric dynamo mode or 
a mixture of two modes and an unusual outward directed orientation 
of the magnetic field (Sect~\ref{s:genuine}). 
According to the magnetic maps, the total magnetic field is stronger 
in NGC\,4254 than in a typical field galaxy (Fig.~\ref{f:magmaps}, Table~\ref{t:mf}). 
Such higher magnetic activity can be associated with the high SFR  
manifested by this galaxy (Paper I). However, several magnetic arms with strong 
regular fields located on the downstream and upstream sides of local density waves 
cannot be readily accounted for and different physical processes 
could be involved. The strong regular field (of about $12\,\mu$G) revealed 
not only in 
the southern galactic outskirts, but also in the western outer magnetic arm, 
well inside the galactic disk, suggests a possibility that both the magnetic 
structures could have arisen from the same phenomena. 
To address all these problems, below we analyze the 
correspondence of the orientation of the magnetic field with the optical spiral arms 
and the connection of the regular field with star-forming regions. 
Subsequently, we model depolarization and magnetic field components, according to 
different physical phenomena.

\subsection{Orientations of magnetic field vectors}
\label{s:orient}

We can explore the relation between the magnetic and perturbed optical spiral 
arm pattern in NGC\,4254 by investigating the orientation of magnetic field vectors 
with respect to the main optical spiral arms. We constructed a 
phase diagram of orientations of such vectors, $RM$--corrected and de-projected, 
as a function of radius $R$ (in natural logarithm) and galactocentric
azimuthal angle. 

Throughout large galaxy domains the magnetic field vectors show a similar pitch 
angle (Fig.~\ref{f:orient}). In the inner disk part ($R<50\arcsec$) two magnetic patterns 
are observed. The dominant one has a coherent large pitch angle ($>25\degr$), regardless 
of the chaotic pattern of H$\alpha$ and UV emitting gas. 
This pattern is broken in two regions (around the azimuthal angle of 
$150\degr$ and $340\degr$) with almost azimuthal magnetic field. Small pitch angles 
could be associated with the central bar-like structure seen in high-resolution 
($7\farcs 5$) radio maps (Paper I). Indeed, the position angle of the bar visible 
in high-resolution radio observations (see Fig.~3 in Paper I) 
is $55\degr$ giving the azimuthal angles of the bar ends in the regions of low 
values of the magnetic pitch angle. A similar configuration of magnetic field (with azimuthal 
field close to the bar ends) is visible in the strongly-barred galaxy NGC\,1365 
(Moss et al. \cite{moss07}). However, the performed MHD simulation was unable to 
reproduce this magnetic pattern in detail (the mean difference between 
observed and simulated pitch angles is $30\degr$). In order to fully explain the 
magnetic field within the inner disk of NGC\,4254, an advanced MHD simulation of a weakly 
barred galaxy is needed.

In the southern polarized ridge (radius 
$60\arcsec-100\arcsec$ and azimuth $45\degr-150\degr$), the regular magnetic 
field changes its pitch angle and the polarized emission is shifted outside the 
local UV-maxima. The magnetic vectors seem to be aligned with the optical (and UV) 
SW spiral arm, which has an oscillating pattern around the pitch angle of about $20\degr$ 
(marked in Fig.~\ref{f:orient}). 

At larger azimuth ($>180\degr$), the N and NE optical spiral arms have 
different average pitch angle than the SW arm ($30\degr$ and $25\degr$, respectively), 
providing further evidence for the perturbed spiral pattern of NGC\,4254.
The N optical arm is heavily curved at large distances ($r>30\arcsec$) 
and its pitch angle is reduced to zero. The $\vec{B}$-vectors 
trace the orientation of both the optical arms and deviate from them only slightly 
(up to $20\degr$), even at the tip of the N arm. The polarized maxima are shifted 
from the optical spiral arms, especially on the downstream side of the N arm. In this 
region (between SW and N arms), the magnetic pitch angle reaches a locally large value 
of $50\degr$. We showed in Sect.~\ref{s:magmaps} that in this part of the galaxy 
the energy density of the ISM can be dominated by the magnetic component, thus probably 
allowing the magnetic field to form its own configuration.

The following of the perturbed optical spiral arms by the vectors of 
the regular magnetic field suggests a physical association of both patterns. 
Similar correlations has been observed for some grand-design spirals, 
e.g., in M\,51 (Fletcher et al. \cite{fletcher04b}; Patrikeev et al. 
\cite{patrikeev06}). The alignment of $\vec{B}$-vectors with CO pitch angles in M\,51 was 
suggested to results from the shock compression along the upstream side of 
spiral density waves. The northern part of NGC\,4254 is much more 
flocculent than the spiral structure in M\,51, which implies that the 
suggested interaction must still hold for weak density waves and for a 
perturbed disk. However, the downstream magnetic arms in NGC\,4254 cannot be 
explained by this process.

Locally, large magnetic pitch angle correlated with orientation of spiral 
arms in NGC\,4254 is not attainable in the classical MHD dynamo model (Elstner 
et al. \cite{elstner00}). Relatively-large, magnetic pitch angles (up to 
$40\degr$) could still be obtained in MHD dynamo by allowing an increase in the 
correlation time of interstellar turbulence (Rohde \& Elstner \cite{rohde97}; 
Rohde et al. \cite{rohde99}). However, the magnetic pitch angle was found 
in this case not to depend on the pitch angle of the gaseous arms. 
Furthermore, this approach was criticized by Shukurov (\cite{shukurov05}), 
who argued for shorter correlation time in spiral arms, contrary to the authors above.
MHD simulations by Elstner et al. (\cite{elstner00}), which include an enhancement 
of turbulent diffusion in spiral arms and density wave-alike gas velocity 
patterns, produced the influence of gas flows on dynamo-generated magnetic fields.
However, the generated magnetic pitch angle in the 
interarm regions are smaller than within spiral arms. This is not actually 
observed in NGC\,4254: in the large interarm region between SW and N 
optical arms, the pitch angle is the largest in the whole galaxy (up to $50\degr$). 
Further investigation by MHD modeling to show whether gas flows or 
high turbulent diffusion are able to maintain such a large pitch angle 
in NGC\,4254 is highly needed.

Addressing the problem of observed phase shifts between the magnetic and 
optical arms in NGC\,4254, we mention the work of Shukurov (\cite{shukurov98}), 
who suggested that a similar shift observed in NGC\,6946 can result from a 
certain time lag between an enhancement in turbulence within density waves and 
the response in dynamo coefficients. This cannot account for the outer 
magnetic arms in NGC\,4254; the lag in these cases would be negative.
There is another possibility given by Shukurov explaining  
magnetic arms in NGC\,6946 as a suppression of dynamo efficiency in the 
gaseous arms by an enhanced turbulent 
magnetic diffusivity. However, such a shift becomes smaller along the SW arm
and disappears in the western part of the arm (azimuth $>140\degr$), contrary 
to the expectations, since a larger distance from the galaxy center should result in 
a smaller dynamo number and a higher shift (as observed, e.g., in M\,51, Fletcher 
et al. \cite{fletcher04b}). Hence, the magnetic arms of NGC\,4254 and the large 
magnetic pitch angle indicate a strong influence of gaseous flows on 
the dynamo-induced magnetic fields, and possibly 
some external cause, such as compression or shear, at work (see below).

\subsection{Magnetic field regulation by SFR}
\label{s:sfr}

\begin{figure*} 
\centering
\includegraphics[width=9.1cm]{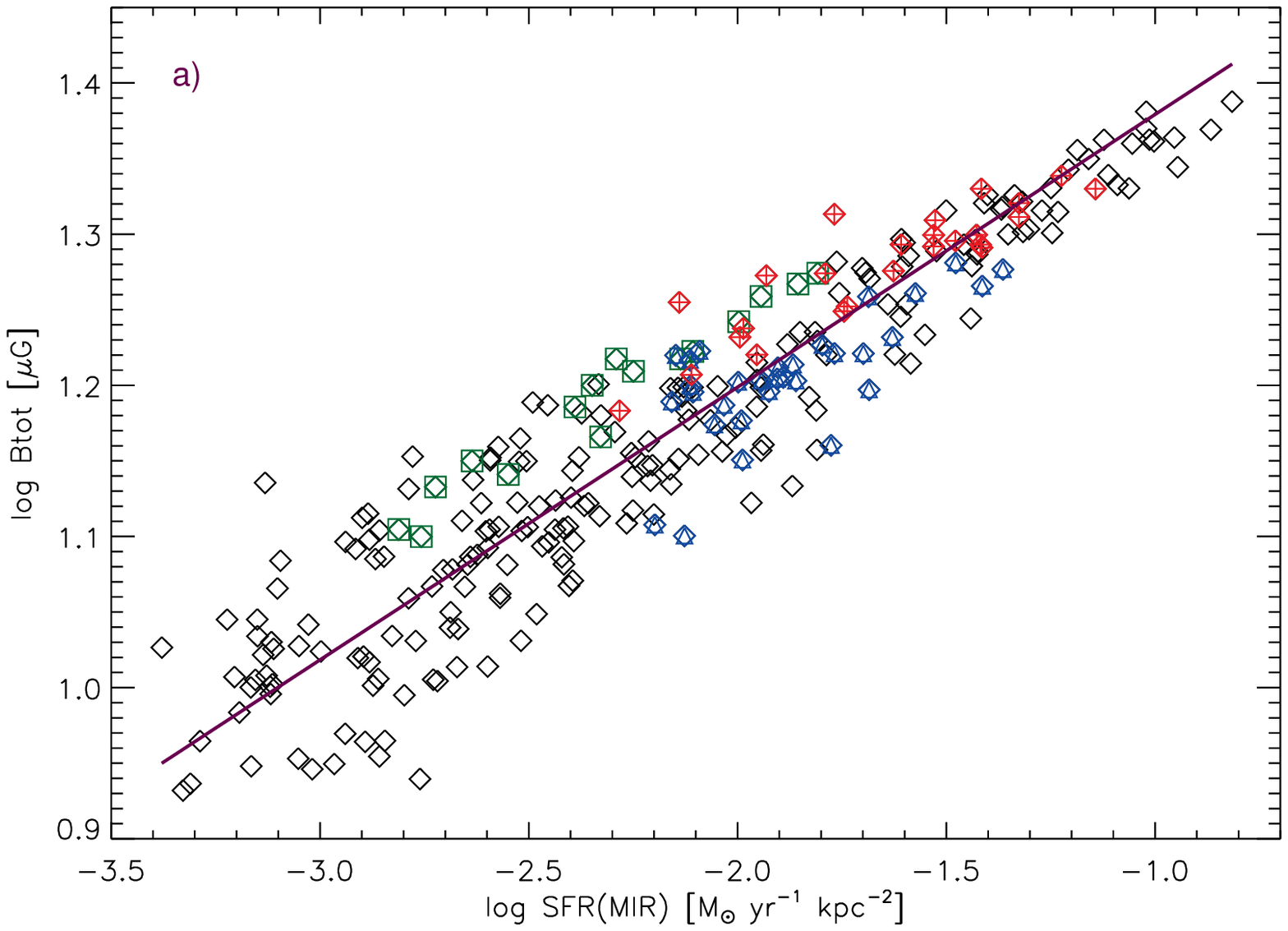}
\includegraphics[width=9.1cm]{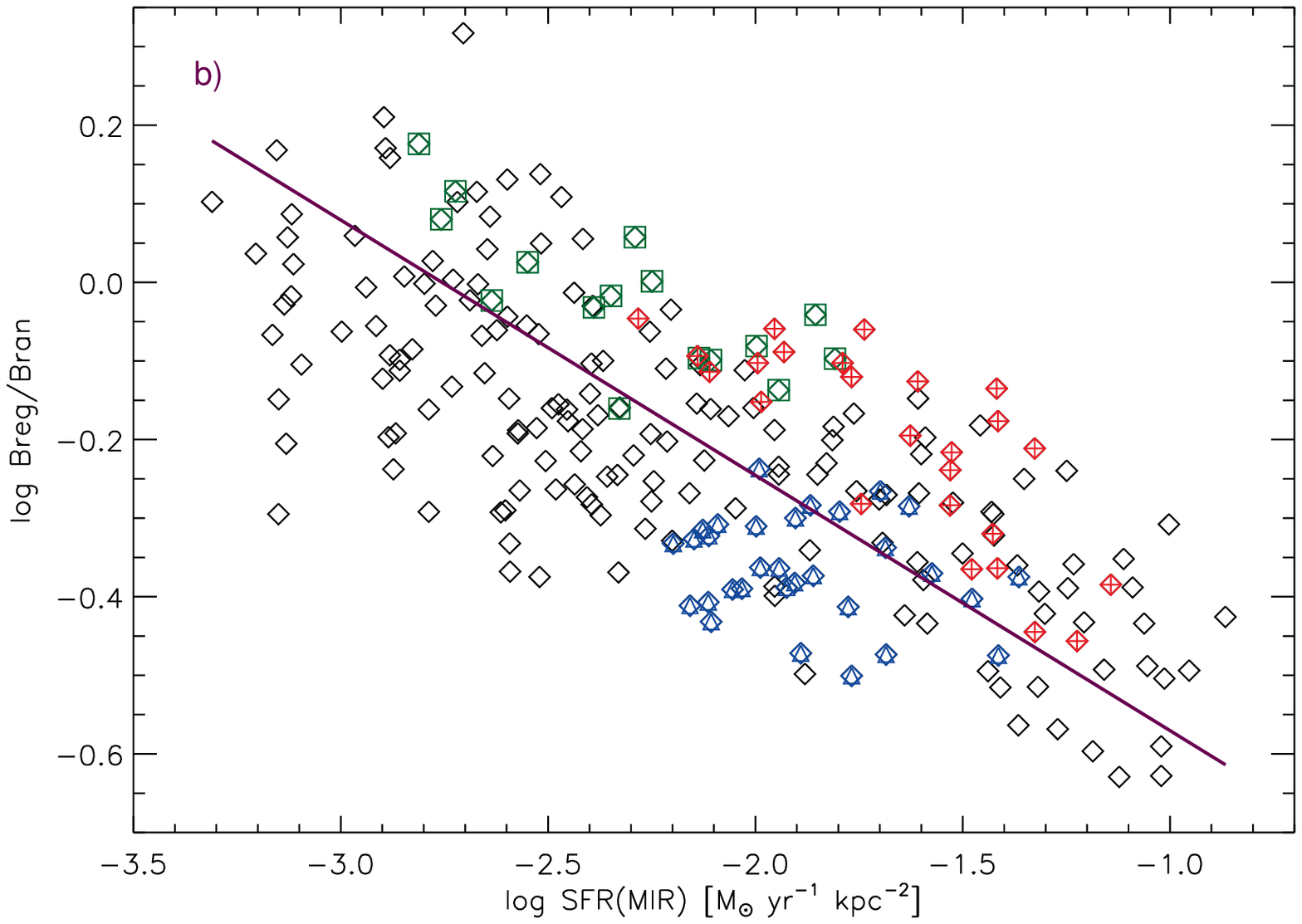}
\includegraphics[width=9.1cm]{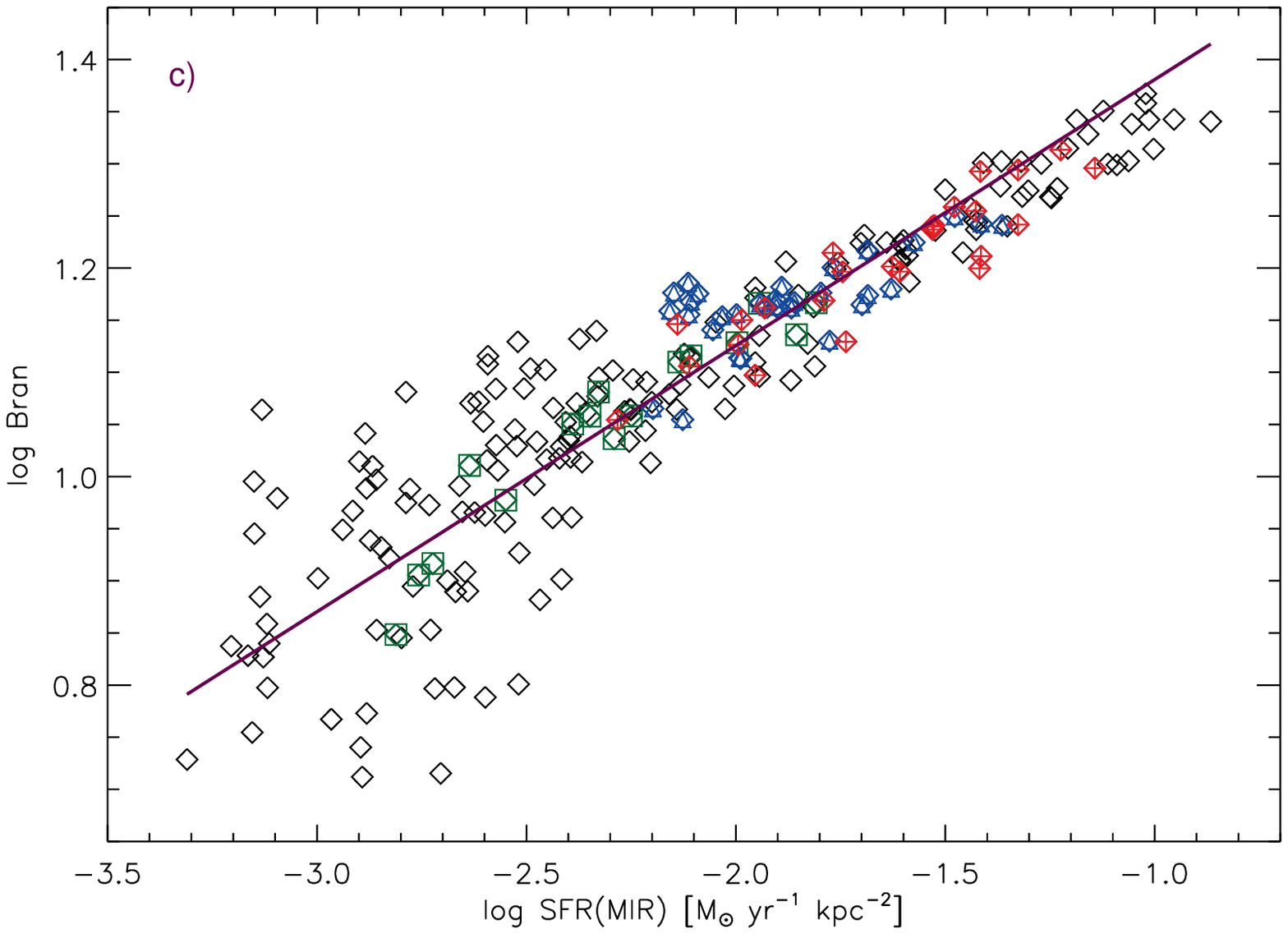}
\includegraphics[width=9.1cm]{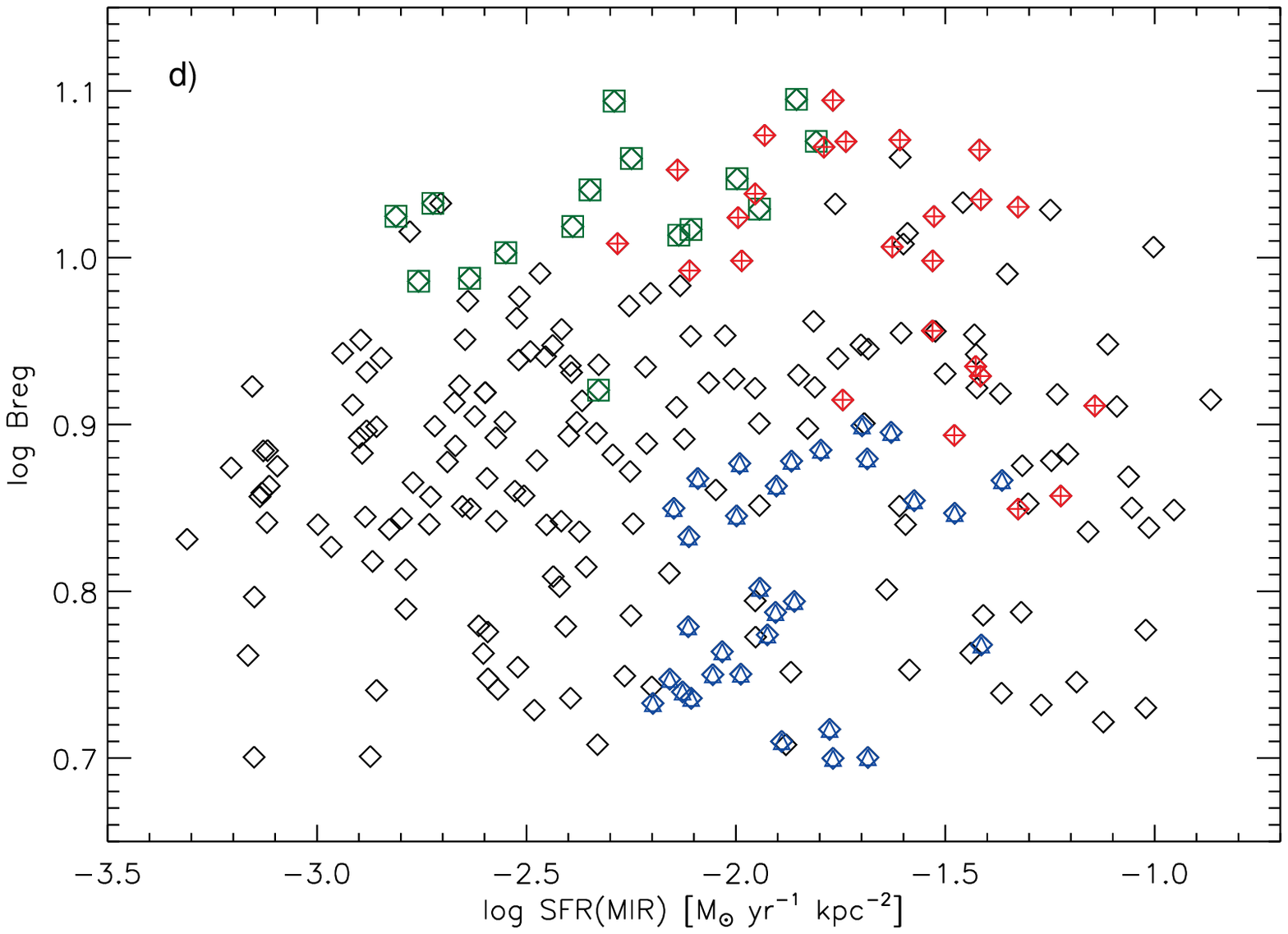}
\caption{
Total magnetic field strength (a), field regularity (b), random 
(c), and regular (d) magnetic field strengths versus 
IR-based SFR within NGC\,4254. From among all regions (marked by 
diamond sign) magnetic arms are denoted separately: southern outer 
magnetic arm by crosses; western outer magnetic arm by 
rectangles; and northern inner magnetic arm by triangles. 
The lines fitted by bisector method to all regions are drawn (except 
d-panel, see Sect.~\ref{s:sfr}).
} 
\label{f:btsfr}
\end{figure*}

Magnetic maps enable analyzing the correlation between the strength of  
various components of the magnetic field and the star formation activity over 
different regions of 
NGC\,4254. In Fig.~\ref{f:btsfr}a, we present such a relation as based on 
24\,$\mu$m SFR (cf. Paper I) and the magnetic map of total field strength 
(Fig.~\ref{f:magmaps}a). The observed relation for 277 beam independent 
($15\arcsec$ spaced) regions over the galaxy is a single power-law 
throughout almost three orders of magnitude in SFR. The relation is tight, 
with the Pearson correlation coefficient of 0.93, and has the 
form fitted by the bisector method:
\begin{equation}
 \log B_{\rm{tot}}=0.18\,(\pm 0.01) \log {\rm SFR}+1.56\,(\pm 0.01).
\end{equation}
This relation is also another example of the well-known radio/IR 
correlation discussed in Paper I and reflects a process of scaling the 
production of magnetic field and CRs with star formation activity. 

We also separated regions of S, W, and N magnetic arms and marked them all 
in Fig.~\ref{f:btsfr}. We repeated the fitting procedures for them. The 
resulting slopes $a$, offsets $b$, correlation coefficients $r$, 
and number of used regions $N$ are given in Table~\ref{t:slopes}. The 
regions of outer magnetic maps have relatively-high total field strength 
as compared to the inner N magnetic arm.

The intensive star formation is generally considered to produce an increased 
level of turbulence and to influence the regular magnetic field. However, 
this process has not been quantitatively investigated to date. Now, for the 
first time, this can be done with the constructed magnetic maps for 
NGC\,4254. The relation between SFR and the regular field's strength scaled 
by the random field -- i.e., the field regularity -- is presented in 
Fig.~\ref{f:btsfr}b for 234 independent galaxy regions. The relation is well 
described by a power-law that we estimated by the bisector method, which 
yields: 
\begin{equation}
\log \frac{\mathrm{B_{reg}}}{\mathrm{B_{ran}}}=-0.32\,(\pm 0.01) \log {\rm SFR}-0.90\,(\pm0.03).
\end{equation}
It shows that contrary to the total magnetic field strength, the field 
regularity strongly decreases with the star-formation level: the observed 
anticorrelation is $-0.71$. Several processes can explain the observed 
trend: 1) field tangling by some processes related to the star formation; 
2) production of turbulent fields proportional to SFR (by a turbulent dynamo 
or any other turbulent field amplification mechanisms); and 3) production of 
regular field anti-proportional to SFR. The latter can be achieved with the 
mean-field dynamo effects anti-proportional to SFR, e.g., by the correlation 
length of turbulence anti-proportional to SFR, as proposed by Rohde et al. 
(\cite{rohde99}), or by the turbulent diffusivity enhanced by shear, as 
proposed by Moss et al. (\cite{moss01}) and Moss et al. (\cite{moss07}), 
or by the suppression of mean-field dynamo action in spiral arms 
(Shukurov \cite{shukurov05}). As all the above concepts give 
different predictions for scaling of regular and random magnetic field 
with SFR, we present separately the strength of both the magnetic components 
versus SFR in Figs.~\ref{f:btsfr}c-d and the fitted power-laws in 
Table~\ref{t:slopes}. Strikingly enough, there is a strong correlation of 
random field $B_{\rm{ran}}$ with SFR ($r=0.91$ for all regions) and no such 
correlation for regular field ($r=0.08$). This suggests that the effective 
production of random magnetic field in the star-forming regions (i.e., the 
second possibly underlying process from those listed above) is the principal 
cause of the strong anticorrelation of field regularity and SFR 
(Fig.~\ref{f:btsfr}b). 

However, the $B_{\rm{reg}}-$SFR relation shows a two-way behavior 
(Fig.~\ref{f:btsfr}d) for stronger regular fields with an upper envelope 
slightly raising for weak SFR, and declining for the large SFR. The dividing 
value is $\mathrm{SFR}\approx 0.016\,\mathrm{M}_
{\sun}\,\mathrm{yr}^{-1}\,\mathrm{kpc}^{-2}$, which corresponds to 
$B_{\rm ran}\approx 14\,\mu$G and to the maximum of $B_{\rm reg}$ of about 
$12.5\,\mu$G. This two-way trend is apparent in the data in 
Table~\ref{t:slopes} as positive and negative correlation coefficients for 
western and southern magnetic arms, respectively, and an insignificant 
correlation when all galactic regions are taken together. 
Hence, the production of regular field in NGC\,4254 
could have been suppressed {\em in the most vivid star-forming regions} (process 3) 
or the regular field could have been disrupted by an efficient field tangling 
(process 1 above). Further analyses of magnetic structures in other galaxies 
and advanced simulations of MHD dynamo involving various SFRs are needed to 
fully account for these relations. 

\begin{table}
\caption[]{The fitted relations between total $B_{\rm{tot}}$, 
random $B_{\rm{ran}}$, and regular $B_{\rm{reg}}$ magnetic 
field strength, and the field regularity 
${B_{\mathrm{reg}}}/{B_{\mathrm{ran}}}$, versus star-formation 
rate SFR for all regions of NGC\,4254 and for different magnetic arms.
}
\label{t:slopes}
\centering
\begin{tabular}{lllll}
\hline\hline
regions &  $a$           & $b$             & $r$   & $N$  \\
\hline
        &\multicolumn{2}{l}{$\log B_{\rm{tot}}=a 
\log \mathrm{SFR} + b$}\\
all     & $0.18\pm0.01$  & $1.56\pm0.01$   & $+0.93$ & 277\\
S arm  & $0.13\pm0.01$  & $1.50\pm0.02$   & $+0.89$ & 25\\
W arm   & $0.17\pm0.01$  & $1.59\pm0.02$   & $+0.98$ & 16\\ 
N arm   & $0.19\pm0.03$  & $1.56\pm0.05$   & $+0.72$ & 31\\ 
\hline
        &\multicolumn{2}{l}{$\log B_{\rm{ran}}=a 
\log \mathrm{SFR} + b$}\\
all     & $0.26\pm0.01$  & $1.64\pm0.02$ & $0.91$ & 234\\
S arm  & $0.22\pm0.02$  & $1.57\pm0.03$ & $0.91$ & 25\\
W arm   & $0.30\pm0.03$  & $1.73\pm0.06$ & $0.95$ & 16\\ 
N arm   & $0.20\pm0.03$  & $1.55\pm0.05$ & $0.72$ & 31\\ 
\hline
        &\multicolumn{2}{l}{$\log B_{\rm{reg}}=a 
\log \mathrm{SFR} + b$}\\
all     &\multicolumn{2}{c}{two-way trend}  & $0.08$ & 234\\
S arm  & $-0.28\pm0.04$  & $0.52\pm0.07$ & $-0.49$ & 25\\
W arm   & $0.17\pm0.07$  & $1.43\pm0.15$ & $0.46$ & 16\\ 
N arm   & $0.47\pm0.16$  & $1.69\pm0.05$ & $0.28$ & 31\\ 
\hline
        &\multicolumn{2}{l}{$\log \frac{B_{\rm{reg}}}{B_{\rm{ran}}}=a 
\log \mathrm{SFR} + b$}\\
all     & $-0.32\pm0.01$  & $-0.90\pm0.03$ & $-0.71$ & 234\\
S arm  & $-0.42\pm0.05$  & $-0.90\pm0.08$ & $-0.78$ & 25\\
W arm   & $-0.30\pm0.04$  & $-0.71\pm0.11$ & $-0.75$ & 16\\ 
N arm   & $-0.56\pm0.21$  & $-1.41\pm0.39$ & $-0.21$ & 31\\ 
\hline
\end{tabular}
\end{table}
The regions of S and W outer magnetic arms reveal a distinct behavior from 
the other galactical regions. Their regular field strength reaches the highest 
attainable values, which places them on the top of $B_{\rm{reg}}-$SFR relation 
(Fig.~\ref{f:btsfr}d). In contrast, their random field is average, 
situating them close to the fitted power-law for all the galactic regions 
(Fig.~\ref{f:btsfr}c). Thus, the stronger total field for 
the outer magnetic maps mentioned above is due to the stronger regular field and not the 
random one. 

This behavior is well rendered by the field regularity data. On the field 
regularity-SFR diagram (Fig.~\ref{f:btsfr}b) the outer magnetic arms approach 
large values at the upper envelope of the relation. In contrast, the regions 
of the inner N magnetic arm are distributed mostly below the average trend 
determined for all the galaxy regions (the fitted line in 
Fig.~\ref{f:btsfr}b). This can indicate that the regular field in the outer 
magnetic arms is probably not fully controlled by the star formation, as for 
instance in the southern outer magnetic arm, which shows a high-field regularity 
and does not avoid the high-SFR regions. Apparently, some additional process 
to regularize magnetic fields or a more efficient production of regular field 
component must be at work within the outer magnetic arms (which we explore 
in Sect.~\ref{s:ani}).

We also investigate how the applied assumption of constant nonthermal 
spectral index $\alpha_{\rm nth}$ in derivation of magnetic maps 
in Sect.~\ref{s:magmaps} could influence the discussed above relations. 
We use from Sect.~\ref{s:magmaps} a rough estimation of possible 
changes of the total magnetic field in the NE spiral arm and in the 
tip of the western outer magnetic arm. 
We predict a possible slight decrease of the slope in $B_{\rm{tot}}-$SFR 
relation which would, however, give a power-law fit within the current 
spread of points in Figure~\ref{f:btsfr}a. Similar considerations predict 
a possible small decrease of the slope of $B_{\rm{ran}}-$SFR relation 
(Fig.~\ref{f:btsfr}c) and an increase of the anticorrelation in the field 
regularity-SFR relation (Fig.~\ref{f:btsfr}b), again within the spread of 
the data points. The variations of $\alpha_{\rm nth}$ 
across the disk would induce an increase of regular field in galaxy 
outskirts and a decrease in star-forming regions, which we suspect could 
lead to a slight predominance of anticorrelation in the $B_{\rm{reg}}-$SFR 
relation (Fig.~\ref{f:btsfr}d).

\begin{table*}
\caption[]{
Observed and modeled depolarization of synchrotron emission 
in several regions of NGC\,4254 due to differential Faraday rotation, Faraday 
dispersion, and gradients in $RM$. The regions are marked in Fig.~\ref{f:magmaps}b.
}
\begin{center}
\begin{tabular}{lrrrr}
\hline
\hline
                          & obs. $DP$(3.6/6.3) & Diff. FR      & Far. Disp.    & Grad. $RM$  \\
\hline                                                      
1) Southern outer magnetic arm & $1.08\pm0.02$ & $ 0.99\pm0.01$ & $0.99\pm0.01$ & $0.99\pm0.01$ \\
2) Western outer magnetic arm  & $1.14\pm0.14$ & $ 0.97\pm0.02$ & $1.00\pm0.01$ & $0.98\pm0.01$ \\
3) Northern inner magnetic arm & $1.04\pm0.17$ & $ 0.99\pm0.01$ & $1.00\pm0.01$ & $0.98\pm0.01$ \\
4) Central part                & $0.57\pm0.18$ & $ 0.66\pm0.28$ & $0.70\pm0.04$ & $0.91\pm0.11$ \\
5) SW optical spiral arm       & $0.89\pm0.04$ & $ 0.99\pm0.01$ & $0.86\pm0.06$ & $0.99\pm0.01$ \\
6) NE optical spiral arm       & $0.72\pm0.15$ & $ 0.91\pm0.07$ & $0.89\pm0.15$ & $0.94\pm0.04$ \\
\hline \\
\label{t:depo}
\end{tabular}
\end{center}
\end{table*}

\subsection{Modeling depolarization}
\label{s:depo}

To establish the origin of magnetic field in NGC\,4254, and the processes that 
underlie the observed regularity of magnetic field, it is necessary to understand 
the depolarization processes that influence the observed properties of polarized 
emission. In this section, we model various frequency-dependent  
depolarization effects in NGC\,4254 by constructing distributions (maps) of 
expected depolarization and comparing them with observed distribution (Fig.~\ref{f:rm}b).

The line-of-sight component of coherent magnetic field embedded 
within the synchrotron emitting region and mixed with ionized gas can 
cause the differential Faraday rotation (Burn \cite{burn66}). To obtain the 
modeled values of the depolarization in NGC\,4254, we use the formula 
(Sokoloff et al. \cite{sokoloff98}):
\begin{equation}
DP_{dFR}=\left| \frac{\lambda^2_2}{\lambda^2_1}
\frac{\sin(2 RM \lambda^2_{1})}{\sin(2 RM \lambda^2_{2})}\right|
\label{e:dfr}
\end{equation}
where $\lambda_1=0.062$\,m, $\lambda_2=0.035$\,m, and the rotation measure $RM$ 
values we take from the constructed $RM$ map (Fig.~\ref{f:rm}a).

The modeled depolarization, due to this effect, and observed depolarization are 
presented in Table~\ref{t:depo} for some characteristic places within NGC\,4254. 
Uncertainties of depolarization values estimated from their standard deviations in 
the measured regions are also given in Table~\ref{t:depo}. Throughout almost 
the entire galaxy the calculated depolarization is weak, $DP_{dFR}$ values are 
close to about 0.95. A larger modeled depolarization (with large statistical 
variations) appears in the central disk region, where $DP_{dFR} \approx 0.7\pm0.3$, 
and in the regions of $RM$ jumps, where 
$DP_{dFR}<0.2$, and where observed $DP$ are also similarly low. However, variations 
of depolarization are large in those regions. 
The differential Faraday rotation does not account for the observed depolarization 
in the most NE disk portion (around 0.7, Fig.~\ref{f:rm}b), as it typically gives 
values around 0.9. 

As the second process of depolarization, we consider the internal Faraday
dispersion produced by a turbulent magnetic field and thermal gas within 
the emitting volume (Burn \cite{burn66}; Sokoloff et al. \cite{sokoloff98}). 
The fractional depolarization resulting from this process is given by:
\begin{equation}
DP_{disp}=\frac{\lambda^4_{2}}
               {\lambda^4_{1}} 
           \frac{1-\exp(-2\sigma^2_{RM} \lambda^4_{1})} 
                {1-\exp(-2 \sigma^2_{RM} \lambda^4_{2})}. 
\end{equation}
The dispersion of rotation measure $\sigma_{RM}$ can be described using 
the ``random walk'' approach:

\begin{equation}
\sigma_{RM}=0.812 B_r n_c d \sqrt{N}
\end{equation}
where $B_r$ is the component of turbulent magnetic field in the direction of the 
observer, $d$ is the turbulent cell/cloud size (correlation length) of the 
random magnetic field, $n_c$ is the electron density within the cloud, and 
$N$ is the cloud number along the line of sight. If the ionized diffuse gas 
has an extent $L$ toward the observer, and has a filling factor $f$, then its 
mean electron density $<n>=f n_c$ and $N=L f d^{-1}$.
Similar results (with $\sigma_{RM}$ larger by up to two in Eq. 6) are 
obtained using different approach (Beck et al. \cite{beck03}).
 
The magnetic maps of random magnetic field strength allow us to derive a map 
of its line-of-sight component $B_r=B_{\rm ran}/\sqrt{3}$. 
Taking typical values of parameters: $d=50$\,pc, $L=1$\,kpc, and $f=0.5$ 
(cf. Beck \cite{beck07}), 
we can adjust the only free parameter -- the cloud density $n_c$ -- to get 
the best correspondence of the modeled $DP_{disp}$ with the actual map of 
observed depolarization within the spiral arms, where the Faraday dispersion 
is expected to be most pronounced. We find the best value 
$n_c=0.08$\,cm$^{-3}$ which yields $<n>=0.04$\,cm$^{-3}$, close to typical values used 
(e.g. $<n>=0.03$\,cm$^{-3}$ for NGC\,6946; Beck \cite{beck07}). Within these numbers Faraday 
dispersion can almost fully explain the actual low values of depolarization 
in the center of NGC\,4254 giving at the same time good correspondence 
within the magnetic arms (Table~\ref{t:depo}). In the most NE part of the galaxy 
with rather weak thermal emission the obtained Faraday dispersion 
($DP_{disp}>0.95$) cannot account for the observed values (about 0.7).

We can also obtain the mean electron density from the emission measure $EM$ of 
diffuse thermal gas: $<n>=\sqrt{f<n^2>}=\sqrt{f EM L^{-1}}$ which yields 
\begin{equation}
\sigma_{RM}=0.812 B_r \sqrt{EM d},
\end{equation}
independent of the filling factor. 
The emission measure $EM$ can be estimated from our map of radio thermal 
emission of NGC\,4254 at 8.46\,GHz. 
Following Walterbos (\cite{walterbos00}), we assume that in NGC\,4254, 
as in the Galaxy and in M\,31, about 20\% of the total ionized 
gas is diffuse, and that only this ISM phase contributes to $RM$ and 
to dispersion of $RM$. To make the modeled depolarization map 
consistent with observations, we must assume a 
correlation size $d$ in Eq.~(7) of about 0.5\,pc. However, this value is much 
smaller than applied above in Eq.~(6), and which is usually assumed 
for other galaxies (see e.g., Fletcher et al. \cite{fletcher04a}; 
Beck \cite{beck07}). For regions of high $EM$ ($>1000$\,pc\,cm$^{-3}$) 
it also evokes unreasonable small filling factor of the diffuse gas 
$f$, below 0.04. Either the used assumption in Eq.~(7) of 20\% diffuse 
gas is wrong, which would mean that NGC\,4254 is a quite different 
galaxy from the Milky Way and M\,31 due to, e.g., a much higher SFR, 
or our depolarization model is insufficient. An independent way to
determine locally the amount of the diffused ionized emission in 
NGC\,4254 is needed to solve this puzzle. Hence, fully understanding 
Faraday depolarization effects (as modeled by Eqs. 6 and 7) seems 
essential in building a consistent model of the ISM in galaxies. 

The third process, we will explore is depolarization $DP_{gFR}$ by 
gradients in the $RM$ across the observed synthesized beam of $\theta$ width, 
which occurs in a foreground screen (Sokoloff et al. \cite{sokoloff98}). 
We consider linear $RM$ gradients that are resolved with 
the observed beam. If the $RM$ has unresolved fine structure (within the 
beam size) then depolarization could be larger. Hence, in our approach 
the calculated $DP_{gFR}$ values represent an upper limit of depolarization. 
We constructed a relevant depolarization model map for NGC\,4254 using 
AIPS task BDEPO, which use the following approximation:
\begin{equation}
DP_{gFR}=exp\left[\frac{-(g_{RM} \theta)^2 (\lambda^4_1 - \lambda^4_2)}{2\ln 2}\right]
\end{equation} 
where $g_{RM}$ is the gradient in $RM$. We estimate this gradient by applying 
the Sobol operator to the observed $RM$ map with the help of the AIPS 
task NINER. 

The modeled $DP_{gFR}$ values outside the core of NGC\,4254 are typically 
larger than 0.95 (Table~\ref{t:depo}). Relatively low values of 0.75-0.85 are predicted
for the NE outskirts of the galaxy, where observed depolarization is at a similar 
level and which could not be accounted for by other depolarization effects. As expected, 
gradients in $RM$ cause a large amount of depolarization in the $RM$ jump regions, where 
$DP_{gRM}$ goes down to 0.2-0.4.

In summary, various depolarization processes dominate in different regions of 
NGC\,4254, which explain the conclusion of Sect.~\ref{s:rm}, where we found no global 
correlation of depolarization with $RM$ or thermal emission. 
The depolarization observed in the central part of the galaxy is well accounted 
for by Faraday dispersion together with differential Faraday rotation 
(Table~\ref{t:depo}), while the large depolarization observed around $RM$ 
jumps can be successfully explained by differential Faraday rotation and 
depolarization due to $RM$ gradients. The gradients in $RM$ can also underlie 
the depolarization observed in the NE part of the galaxy. 

None of the modeled processes predict an enhanced level of depolarization 
in the regions of both outer magnetic arms in agreement with observations. 
In fact, a completely different set of data at 1.43\,GHz shows that all 
those regions are still polarized at this frequency (Fig.~4 in Paper I) and 
must be Faraday-thin, thus avoiding strong Faraday effects.
As both outer magnetic arms manifest depolarization properties similar to the 
N inner magnetic arm, they cannot arise from reduced depolarization. 
Therefore, outer magnetic arms require an additional process 
that enhances regular component of the magnetic field.

\subsection{Modeling of magnetic field components}
\label{s:ani}

In Sect.~\ref{s:sfr}, we demonstrated that the outer magnetic arms in NGC\,4254 
involve a statistically more regular magnetic field than the other 
galaxy regions with a comparable level of star-forming activity. We keep in 
mind that the 
regular field strength derived from the polarized intensity (Sect.~\ref{s:rm}) 
contains coherent and anisotropic components.  
Gas compression and stretching (with shearing) are well-recognized processes 
that can modify the coherent field and produce anisotropic field from 
isotropic (random) one (Sokoloff et al. \cite{sokoloff98}, Beck et al. 
\cite{beck05}). In Paper I, we suggested that the southern polarized ridge may 
indeed be affected by stretching/shearing forces of tidal origin. As less 
likely, but still possible, we considered compressional forces acting in 
the southern disk portion due to the ram pressure of hot cluster gas. 
We propose here an analytical model to discriminate between stretching and 
ram-pressure alternatives as well as to explain the origin of outer magnetic arms.

In this model, we describe the galactic magnetic field in the cylindrical polar 
coordinates ($R$, $\phi$, $z$) with ($R$, $\phi$) aligned with the galaxy 
plane and with azimuthal angle $\phi$ counted counterclockwise from the 
northern edge of the galaxy major axis. The observer (sky) plane is inclined 
to the galactic plane by the galaxy inclination angle $i=42\degr$ 
(Paper I). The discussed compression hypothesis 
involves external processes to act only in the outskirts of the galaxy, while 
the western outer magnetic arm, which is inside the galaxy disk, would remain 
unexplained. Therefore, we concentrate on modeling the southern magnetic arm. 

As the initial properties of the modeled southern arm (without stretching or 
compression) we take the magnetic properties of the N arm (Table~\ref{t:mf}),
which we expect to be undistorted by tidal (stretching) or ram-pressure (compression) 
forces and to contain no anisotropic field. We notice that in the N arm (exactly 
region 3, Table~\ref{t:mf}) magnetic field vectors are aligned very close with the 
direction of the galaxy major axis (the difference is about $10\degr$, Fig.~\ref{f:true}). 
In the southern magnetic arm (region 1, Table~\ref{t:mf}) they are oriented almost perpendicularly 
to the galaxy minor axis (difference of about $12\degr$). 
Therefore, without any correction for projection effects we can approximate 
the coherent field in the southern arm by its azimuthal component only 
of strength equal to the regular field observed in the N arm.
Hence, following the approach of Sokoloff et al. (\cite{sokoloff98}) the 
components of magnetic field $\vec B_1$ in the modeled southern 
arm are initially: 
\begin{eqnarray}
\nonumber B_{\phi,1}&=&\bar{B}+b_{\phi,1}\\
B_{R,1}&=&b_{R,1} \\
\nonumber B_{z,1}&=&b_{z,1}
\end{eqnarray}
where $\bar{B}$ denotes volume (ensemble) averaged magnetic field, thus its 
coherent part $\bar{B}=\left<B_{\phi,1}\right>$ and $b_{\phi,1}$, $b_{R,1}$, 
$b_{z,1}$ are random magnetic field components, with $\left<b_{\phi,1}\right> = 
\left< b_{R,1}\right> = \left<b_{z,1}\right> = 0$. Assuming at the beginning 
a fully isotropic random field, we demand $\left<b^2_{\phi,1}\right>=
\left< b^2_{R,1}\right>=\left<b^2_{z,1}\right>=\sigma^2$, where $\sigma$ is 
the standard deviation of a random field in one dimension. 

We assume that the stretching/shearing is along the southern magnetic arm. This should 
keep constant the strength of the initial coherent field and produce an excess 
in the form of azimuthal random component in the azimuthal direction, thus providing for 
the field anisotropy. We approximately scale the strength of 
this process by a single {\em stretching factor} $\eta$, giving the field 
components in response to stretching in the form:
\begin{eqnarray}
\nonumber B_{\phi,2}&=&\bar{B}+\eta b_{\phi,1} \\
B_{R,2}&=&b_{R,1}\\
\nonumber B_{z,2}&=&b_{z,1}
\end{eqnarray}
In our approach, $\eta=1$ means no stretching/shearing, which corresponds to 
the shear strength $K=0$ in the work of Beck at al. (\cite{beck05}).

With these magnetic field strengths, we can derive two non-dimensional quantities 
in the observer frame: the ratio $S_2/S_1$ of the observed total synchrotron intensities 
from the southern arm after ($S_2$) and before stretching ($S_1$); and the degree of 
polarization $p_2$ of synchrotron emission after stretching. In 
these derivations, we generally follow the approach of Sokoloff et al. 
(\cite{sokoloff98}) for the case of magnetic field and cosmic ray energy 
equipartition. In this case, the number density of CRs $N_{\rm CR}$ scales 
with the magnetic field as $B^2$. However, due to the CRs diffusion, we do 
not expect a complete spatial correlation between both quantities 
to hold locally,  i.e., on scales below 1\,kpc (cf. discussion in Beck at al. 
\cite{beck03}). We then treat both the quantities as locally 
statistically independent variables whose ensemble 
averages satisfy $\left< N_{CR}\right>\propto \left< B^2\right>$. 
Under this condition we get:
\begin{equation}
\frac{S_2}{S_1}=\frac{\left< B^2_2 \right> \left< B^2_{\perp,2} \right> }
{\left< B^2_1 \right> \left< B^2_{\perp,1} \right> }
=\frac{\left[ \bar{B}^2+\sigma^2 \left( \eta^2+2\right) \right] \left[ 
\bar{B}^2+ \sigma^2 \left( \eta^2+1\right) \right]}{\left( \bar{B}^2+3\sigma^2 \right) \left( \bar{B}^2+2\sigma^2 
\right)}
\end{equation}
where $B_{\perp,1}$ and $B_{\perp,2}$ are perpendicular to the line of sight 
components of the magnetic field in the southern arm before and after stretching, 
respectively, and $B_1$ and $B_2$ are the corresponding total field strengths.

\begin{table}
\caption[]{
Modeling of the southern magnetic arm of NGC\,4254 by stretching/shearing and 
compression processes. The radio properties of the N magnetic arm are 
regarded as the initial conditions for the S magnetic arm (region 3 and 1 in 
Table~\ref{t:mf}, respectively). The compression and stretching factors were 
adjusted so as to yield the observational values (given in italics) of 
synchrotron emissivity ratio $S_{2}/S_2$ or the degree of polarization in 
the southern arm $p_{2}$ (see Sect.~\ref{s:ani}). 
}
\begin{center}
\begin{tabular}{lccc}
\hline
\hline
               &  stretching $\eta$ or   & $S_{2}/S_1$ & $p_{2}$  \\
               & compression $\zeta$ factor &                       \\
\hline
Observed     &   --      & $2.0\pm 0.2$     & $0.37\pm 0.03$ \\
\hline
Stretching             &   1.53   & \em{2.0} & 0.39 \\
Stretching             &   1.47    & 1.8     & \em{0.37} \\
\hline
Compression            &   1.24   & {\em 2.0}      & 0.29 \\
Compression            &   1.84   & 8.1     & {\em 0.37} \\
\hline \\
\label{t:compr}
\end{tabular}
\end{center}
\end{table}
For the degree of polarization after stretching we obtain:
\begin{equation}
p_2=p_0\frac{\bar{B}^2+\left( \eta^2-1\right) \sigma^2 }{ \bar{B}^2+\left( 
\eta^2+1 \right) \sigma^2}
\end{equation}
where $p_0\approx 0.75$ is the maximum degree of polarization. We notice that 
$S_2/S_1$ as well as $p_2$ do not depend on the galaxy inclination, as expected 
for the orientation of magnetic vectors in the southern arm, perpendicular to the minor 
axis.

In our modeling of the southern arm, we adjust the stretching factor $\eta$ to get 
exactly the observed value of the ratio $S_2/S_1$ (given in Table~\ref{t:compr}) 
and then compare the modeled $p_2$ to the observed value (Table~\ref{t:compr}). 
In the second approach, we adjust $\eta$ to get for $p_2$ its observational value 
and then we compare modeled $S_2/S_1$ to the actual value. The results 
presented in Table~\ref{t:compr} show that the modeled values fit very well 
the observed ones within their uncertainties. So a tidal stretching can 
very effectively produce the anisotropic magnetic field in NGC\,4254. About 
60\% of the observed polarized intensity of the southern magnetic arm 
comes in this case from the anisotropic random field. The determined 
stretching factor $\eta\approx1.5$ is rather small when compared, e.g., to the 
typical shear strengths in strongly-barred galaxies (about 10 in NGC\,1097, 
Beck et al. \cite{beck05}). 

The second process -- compression -- would produce enhanced magnetic field 
components tangent to the compression plane, leaving the component 
normal to it unchanged (see Beck et al. \cite{beck05}). We suspect that in the case of 
NGC\,4254, compression might act in the plane perpendicular to the galaxy disk, 
from the southeastern side of it. We assume that due to compression the 
tangent regular and random 
field components are $\zeta$ times stronger than the original ones. This {\em 
compression factor} can be interpreted as the ratio of gas densities 
($\zeta=\rho_2/\rho_2$) before ($\rho_1$) and after compression ($\rho_2$). 
Hence, the magnetic field in the southern arm after compression is:
\begin{eqnarray}
\nonumber B_{\phi,2}&=&\zeta \,\bar{B}+\zeta\, b_{\phi,1} \\
B_{R,2}&=&b_{R,1}\\
\nonumber B_{z,2}&=&\zeta \, b_{z,1}
\end{eqnarray}
Similar to the analysis of stretching, we derive in case of compression, the 
synchrotron emission ratio:
\begin{eqnarray}
\frac{S_2}{S_1}&=&\frac{\left< B^2_2 \right> \left< B^2_{\perp,2} \right> }
{\left< B^2_1 \right> \left< B^2_{\perp,1} \right> }
=\frac{\left[ \zeta^2 \bar{B}^2+\sigma^2 \left( 2\zeta^2+1\right) \right]}
{\left( \bar{B}^2+3\sigma^2 \right)} \times \\ 
\nonumber &\times & \frac{\left\{ \zeta^2 \bar{B}^2+\sigma^2 \left[\zeta^2 \left(1+\sin^2 i \right) 
+\cos^2 i \right] \right\}}{ \left( \bar{B}^2+2\sigma^2 \right)}
\end{eqnarray}
and the degree of polarization in the southern arm:
\begin{equation}
p_2=p_0\frac{\zeta^2 \bar{B}^2+\sigma^2 \left( \zeta^2 -1\right) \cos^2 i }
{\zeta^2 \bar{B}^2+\sigma^2 \left[ \zeta^2 \left(1+\sin^2 i\right) +\cos^2 i\right] }.
\end{equation}

According to the above expressions, we model that $\zeta\approx 1.2$ gives a 
synchrotron emissivity ratio $S_2/S_1$ equal to the observed value 
(Table~\ref{t:compr}), but the corresponding degree of polarization 
of 0.29 is lower than the observed one ($0.37\pm0.03$). A higher 
compression value, which yields a stronger 
anisotropic field and a sufficient degree of polarization, results in a 
synchrotron emission ratio (8.1) much higher than the one actually observed 
($2.0\pm0.2$). In comparison to 
the stretching process, for the same production of anisotropic field, 
compression produces a larger amount of isotropic random field, resulting 
in a lower degree of the modeled polarization, which is not consistent with observations. 

We checked that our conclusions are not too sensitive to the galaxy orientation: 
lowering the galaxy inclination by 2 degrees (to $i=40\degr$) enhances the 
degree of polarization by only 1\%. 
We also performed an analysis of 
compression in two other cases: the equipartition assumption with a full local 
spatial correlation between the density of CRs and the magnetic field energy, and 
the case of no equipartition, i.e., constant density of CRs. 
Although we regard these assumptions as less probable than those actually adopted 
(no local correlation), we mention that in the case of constant density 
of CRs for the value of $\zeta=1.5$, the synchrotron emission ratio $S_2/S_1$ 
attains the observed value and the degree of polarization reaches $p_2=0.33$, 
which is quite consistent with the observed $p_2=0.37\pm0.03$. However, 
the higher compression amount ($\zeta=1.8$), while yielding a sufficient degree 
of polarization, results in the ratio $S_2/S_1=2.9$, inconsistent with 
observations. The case of equipartition with local spatial correlation 
gives similar results. The reason why different assumptions on equipartition 
(with and without local correlation) give close results in model fitting 
is the low factor $\zeta$ that 
is needed to explain the enhancement of the magnetic field in the southern arm 
of NGC\,4254. Due to nonlinear influence of $\zeta$ on synchrotron emission 
ratio (Eq. 14), much larger differences in results are expected for 
shocks with larger $\zeta$. Such cases (with full local spatial correlation) 
were studied by Beck et al. (\cite{beck05}), see their Fig.~21 for a rough comparison. 

We conclude that the stretching hypothesis does likely provide a proper 
explanation for the high-field regularity in the southern magnetic arm.

\subsection{The influence of the cluster environments}
\label{s:cluster}

One of the most characteristic features of magnetic structure in NGC\,4254 is 
the well-aligned magnetic field in the two outer (southern and western) magnetic 
arms. Although, on the whole, the field regularity seems to be dominated by 
the production of random fields in star-forming regions (Sect.~\ref{s:sfr}), in 
outer magnetic arms the regularity attains larger values than in other disk regions of 
similar SFR, and thus of similar random fields. 

According to our analysis of depolarization (Sect.~\ref{s:depo}), 
the outer magnetic arms are marginally depolarized, likewise the N 
arm, and their high magnetic regularity cannot result 
from a reduced depolarization. This conclusion corresponds with 
similar results on magnetic arms in NGC\,6946 (Beck \cite{beck07}). 
However, in the case of NGC\,4254, two strong (outer) magnetic arms are 
located downstream of nearby density waves, which was not observed in 
other galaxies. Even the northern part of NGC\,4254 shows (inner) magnetic arms 
on the upstream side of spiral arms. It appears that the unusual properties 
of outer magnetic arms in NGC\,4254 are instead associated with some external 
forces and the cluster environment.

Using the modeling of magnetic field components in the southern magnetic 
arm (Sect.~\ref{s:ani}), we have shown that the compression due to the 
external ram-pressure of hot cluster gas does not provide a sufficiently-high 
magnetic field regularity for the observed level of polarized emission. 
In Paper I, we also enumerated other difficulties of this idea, e.g., as 
low depletion of \ion{H}{i} gas and globally-enhanced, instead of 
spatially-truncated, star formation in NGC\,4254. 

There are known examples of galaxies with a distinct action of strong 
compressional forces and enhanced magnetic fields. Such forces were 
argued to be present in the barred galaxies (Beck et al. \cite{beck05}), and 
assessed by the density ratio of pre- and post-shock gas. We searched for such  
a compressional evidence in the contrast ratio of polarized, nonthermal, 
and \ion{H}{i} emission in NGC\,4254, measured between magnetic or optical 
arms and their outside locations separated by 1.5 beams. We inspected five 
such positions indicated in Fig.~\ref{f:intro}, and present the results in 
Table~\ref{t:contrast}. The polarized ridge (locations a and b) and the W 
outer magnetic arm (c) do not reveal any distinctly large change in the 
analyzed distributions as compared with the other regions. 

\begin{table}[t]
\caption[]{
Contrast ratio in and out of the spiral arms in NGC\,4254 in polarized and 
total synchrotron as well as \ion{H}{i} emission. The positions of the inspected 
five regions are denoted in Fig.~\ref{f:intro}.
}
\begin{center}
\begin{tabular}{lrrrrr}
\hline
\hline
                       & a  &  b &  c  &  d  &  e  \\
\hline
Polarized intensity    & 5.3 & 6.1 & 12.9 & 1.9 & 16.6  \\
Nonthermal intensity   & 3.7 & 4.2 &  1.8 & 2.7 & 3.3   \\
\ion{H}{i} emission            & 2.1 & 1.8 &  2.3 & 2.5 & 1.4   \\
\hline \\
\label{contrast}
\end{tabular}
\end{center}
\label{t:contrast}
\end{table}

In view of the data currently available and the performed modeling of magnetic 
field components (Sect.~\ref{s:ani}), we suggest that the most likely hypothesis 
for the enhanced regular field in the outer magnetic arms are stretching forces of gravitational 
(tidal) origin. This prediction can be further checked by $RM$ data, not sensitive to 
anisotropic field. The $RM$ analysis cannot be performed for the southern outer 
magnetic arm (region 1), since the regular field component lies almost in the 
plane of the sky. However, as the strengths of different components of 
magnetic field for both the outer arms are very similar (Table~\ref{t:mf}), 
we suspect that both the arms are enriched in an anisotropic field of similar 
magnitude and origin. Hence, we perform $RM$ analysis for the western magnetic 
arm. The line-of-sight component of coherent magnetic field 
$<B_{coh||}>$ can be estimated from the actual value of rotation measure:
\begin{equation}
RM=0.812 <B_{\rm coh||}> N_e L\,.
\label{e:rm}
\end{equation}
The mean thermal electron density $N_e$ over the galaxy thermal disk pathlength 
$L$ can be estimated from the thermal radio emission (the emission measure 
$EM$), the fraction of the diffuse gas in the total ionized gas (see 
Sect.\ref{s:depo}), and the filling factor $f$. The inner N magnetic arm 
(region 3, Fig.~\ref{f:magmaps}b) has the observed ${RM}=-43\pm17\,
{\rm rad}\,{\rm m}^{-2}$. Supposing for N and W magnetic arms the same 
values for $L$, $f$ and for the fraction of the diffuse gas, we get from (16) 
a prediction of rotation measure in the W arm (region 2) of 
about 150\,rad\,m$^{-2}$, assuming that the entire regular field in that region 
(12.3\,$\mu$G) is coherent. This is much more than the observed 
$+40\pm17\,\rm{rad}\,\rm{m}^{-2}$. Even taking into account 
large statistical uncertainties in the measured $RM$ values, this 
disagreement can suggest that the polarized emission in the outer western 
arm comes not only from the regular coherent field traced by $RM$ data, but to 
a large extent from an anisotropic field, in agreement with the modeling of stretching effect. 

The stretching and shearing origin of the high magnetic regularity in outer magnetic arms 
is also in accord with the study of magnetic field 
orientations (Sect.~\ref{s:orient}) indicating a strong influence of gaseous 
streamlines on the entire magnetic structure in 
that galaxy. In the Antennae merging system, where such stretching forces are 
expected in the NE portion of NGC\,4038, at the base of the tidal tail, there 
is indeed a polarized ridge (Chy\.zy \& Beck \cite{chyzy04}) of high field 
regularity and with magnetic field vectors aligned with local gas flows. 

Recently, Kantharia et al. (\cite{kantharia08}) detected the radio 
envelope of NGC\,4254 at frequencies down to 240\,MHz. They argue that 
the envelope is a signature of the ram-pressure gas, stripped from the 
galactic disk and now expanding. However, 
radio envelopes were also observed around non-cluster spirals (cf. Paper I) 
and their existence cannot be used in favor of ram-pressure effects. The 
argument that the envelope has a steep synchrotron spectrum 
($\alpha_{\rm nth}<-1.0$) is not convincing either, as 
field galaxies also show spectral steepening at the disk outskirts, which 
results from CRs diffusion and energy losses 
(Sect.~\ref{s:sfr}). Besides, comparing the ram pressure with the magnetic 
field pressure, they assumed several times lower magnetic field strength 
($3\,\mu$G) than is actually observed (Fig.~\ref{f:magmaps}a). Next, 
their argument that the high star-formation in NGC\,4254  
is due to ICM wind is surprising as the Virgo Cluster spirals show the 
opposite effect: a lower star formation for galaxies that experienced 
ICM-ISM stripping events (Paper I, Koopmann \& Kenney \cite{koopmann04}). 

Although NGC\,4254 seems to be a 'young' Virgo Cluster member (Paper I), it 
reveals distortions that instead of a ram pressure, they could be 
attributed to a gravitational 
harassment (Moore et al. \cite{moore96}). It is visible in the galaxy's 
perturbed spiral pattern, disturbed \ion{H}{i} emission, and enhanced level 
of star formation. We also propose that the enhanced total and anisotropic 
magnetic field, the strong southern polarized ridge, and the complex magnetic 
field morphology, not found in the field galaxies, could also be the 
manifestation of the harassment process. We expect that the dramatic 
morphological transformations of gas, as well as magnetic field, are still 
ahead of this object, when it is going to pass the Virgo Cluster core, 
and will experience severe ram pressure stripping and global tides. 

Recent observations of other Virgo spirals (Vollmer at al. \cite{vollmer07}, 
We\.zgowiec et al. \cite{wezgowiec07}) show that asymmetrical polarized emission 
are common among Virgo Cluster galaxies. The modeling of magnetic field 
components as applied to NGC\,4254 in Sect.~\ref{s:ani} can also be performed 
for those galaxies. This could at last allow for a statistical analysis of 
diverse influence of ICM on magnetic field structures in cluster spirals. 

\section{Summary and conclusions}
\label{s:summary}

We present the first detailed investigation of magnetic field, 
Faraday rotation measures, and depolarization of synchrotron emission in a spiral 
galaxy -- NGC\,4254 --  embedded in the Virgo Cluster medium. In Paper I, we 
described our radio (VLA and Effelsberg) and X-ray (XMM-Newton) observations and data reduction, 
along with a comprehensive discussion of the galaxy's properties in different wavelengths. 
The complex polarization properties discovered in different regions of the galaxy 
are analyzed here, with the help of the newly developed concept of presenting magnetic 
field in the form of various ``magnetic maps'' (Sect.~\ref{s:magmaps}) and with 
analytical modeling of Faraday effects (Sect.~\ref{s:depo}), as well as modeling of 
compression and stretching forces on different magnetic field components (Sect.~\ref{s:ani}).

The performed study enabled us to obtain the following results:
\begin{itemize}

\item[-] The distribution of $RM$ in NGC\,4254 reveals large areas of coherent 
(unidirectional) magnetic field, which resemble a 
perturbed axisymmetric dynamo mode or a mixture of axisymmetric 
and bisymmetric modes. The magnetic field vectors are oriented outwards from the galaxy 
disk, contrary to other galaxies.

\item[-] The constructed ``magnetic maps'' show that the total magnetic field 
in NGC\,4254 is stronger than in typical field spirals, reaching 
17\,$\mu$G in the interarm regions, 22\,$\mu$G in spiral arms, and even 
25\,$\mu$G in the disk core (assuming 1\,kpc disk thickness). 
The magnetic energy exceeds the thermal one in almost 
all regions of the galaxy disk. It surpass even the turbulent gas energy in the 
southern polarized ridge, the western interarm region, and at the disk 
outskirts, and may become dynamically important.

\item[-] 
The regular field strength reaches the largest value of 13\,$\mu$G 
and the highest field regularity of about 0.8 in the two (southern and western) 
outer magnetic arms. The inner N magnetic arm involves a weaker regular field 
(8\,$\mu$G) and less field regularity (0.5). While the strongest regular 
field occurs in magnetic arms, the total and random field strengths are at the highest in 
the optical arms. 

\item[-] The Faraday-free pitch angles of magnetic field vectors, which change 
throughout the galaxy from zero (in the polarized ridge) to more than $40\degr$ (in 
the north), always keep their orientation close to the optical filaments 
(within $20\degr$). The dynamo-generated magnetic fields must be 
significantly modified in NGC\,4254 by density waves and gas flows.

\item[-] The distribution of depolarization cannot be easily explained by any 
individual process (Sect.~\ref{s:rm}), and thus various Faraday effects must be at work in 
different galaxy regions. Our modeling demonstrated (Sect.~\ref{s:depo}) the 
dominant role of Faraday dispersion and differential Faraday rotation within the 
spiral arms and in the galaxy center, where the observed depolarization is most 
significant ($DP\approx 0.8-0.6$). None of Faraday depolarization effects influence 
the outer or inner magnetic arms in agreement with the observed 
depolarization there. 

\item[-] The strength of total magnetic field does correlate very well with the 
star-formation rate of individual regions in the galaxy disk (the 
correlation coefficient $r=0.95$), giving a well-fitted power-law relation 
with a slope $+0.18\pm0.01$. However, in the same regions, 
the regularity of magnetic field shows a high level of {\em anticorrelation} 
($r=-0.67$) with the star formation activity, which mainly results from 
efficient production of random field in vivid star-forming regions. For 
the first time, we show that this relation is well approximated by a power-law 
with the slope of $-0.32\pm0.01$. The enhanced field regularity in the outer 
magnetic arms are, however, not fully controlled by the star formation.

\item[-] Our modeling of magnetic field components in magnetic arms 
indicates that the enhanced magnetic field regularity 
in the outer magnetic arms can be produced by stretching and shearing forces, 
likely of tidal origin, which produce an anisotropic magnetic component 
and enhance polarized emission. This is supported by 
the analysis of depolarization and Faraday rotation measures, while being 
in accord with the field regularity-SFR relation. Hence, modeling of 
magnetic field components seems to be a great tool to discriminate between 
different physical phenomena acting in cluster environment. 

\end{itemize}

The validity of the proposed stretching origin of enhanced magnetic field in 
the outer arms of NGC\,4254 could be further explored by some advanced MHD 
modeling, able to discriminate between magnetic field components and 
reproduce the perturbed structure of the galactic spiral arms. This kind of 
modeling of magnetic field components performed for NGC\,4254 can also be 
applied to other Virgo Cluster spirals and would allow for a statistical 
analysis of the magnetic field evolution within galaxies influenced by
the cluster medium. 

\begin{acknowledgements}
I would like to thank Dr Rainer Beck and Prof. Marek Urbanik for helpful 
discussions and stimulating comments, especially on depolarization effects. 
This work was supported by the Polish 
Ministry of Science and Higher Education, grant 2693/H03/2006/31.
\end{acknowledgements}

\end{document}